\documentclass[aps,prd,10pt,superscriptaddress,preprintnumbers,nofootinbib,twocolumn]{revtex4-2}
\usepackage[utf8]{inputenc}
\usepackage{uniinput}
\usepackage{amsfonts,amsmath,mathtools,mathrsfs,hyperref,url,bbm}
\hypersetup{
	colorlinks=true,
	allcolors=blue
}

\usepackage{microtype}
\usepackage{cleveref}
\usepackage{enumitem}
\usepackage{xcolor, transparent}
\usepackage{fourier-orns}
\usepackage{physics} %
\usepackage{subcaption}
\usepackage[commentmarkup=uwave, draft]{changes}
\definechangesauthor[name=SG, color=magenta]{SG}
\definechangesauthor[name=UD, color=olive]{UD}

\newcommand{\km}{k_-}
\newcommand{\kp}{k_+}
\newcommand{\nn}{\nonumber}

\begin{document}

\title{Weak gravity at micron scales from dark bubble cosmology\\ and its cosmological consequences}
	
	\author{Ulf Danielsson~}
	\email{ulf.danielsson@physics.uu.se}
	\affiliation{Institutionen för fysik och astronomi,
		Uppsala Universitet, Box 803, SE-751 08 Uppsala, Sweden}
	\author{Suvendu Giri~}
	\email{suvendu.giri@physics.uu.se}
	\affiliation{Institutionen för fysik och astronomi,
		Uppsala Universitet, Box 803, SE-751 08 Uppsala, Sweden}

	\preprint{UUITP-36/25}

	\begin{abstract}
	\noindent
	The dark bubble model makes a positive cosmological constant natural in string theory, and predicts several new physical phenomena within reach in the near future. In this paper we study the experimental consequences of the model for the strength of gravity at scales of order $10^{-5}$m. Contrary to other models of gravity involving extra dimensions, the dark bubble model predicts gravity to become weaker rather than stronger at small scales, compared to Newtonian gravity. In particular, we provide explicit predictions of measurable deviations using table top experiments. We also show how the same effect reduces the effective force of gravity at high energy densities in cosmology, leading to a period of early inflation without the need for anything beyond radiation. We also discuss the quantum origin of the universe with a 5D black hole acting as a catalyst for the nucleation of the dark bubble and how it accounts for the present matter content in the universe. This leads to a prediction of $\Omega_c \approx 5\times 10^{-4}$ for a positive curvature of the universe, suggesting an explanation of the why-now-problem of the cosmological constant. We end by speculating on how to incorporate AdS black shells as black hole mimickers within the dark bubble model.
	\end{abstract}
	
	\maketitle
	\tableofcontents

\allowdisplaybreaks

\section{Introduction}

The dark bubble model \cite{Banerjee:2018qey} proposes that our universe is an expanding bubble embedded into 5D AdS space. Its presence is a consequence of a phase transition in 5D where a bubble of true vacuum nucleates. Its accelerated expansion is understood as the consequence of a positive cosmological constant from the point of view of 4D. Many different aspects of the dark bubble model have been investigated over the last several years (see \cite{Banerjee:2021yrb,Danielsson:2022fhd,Danielsson:2022lsl,Basile:2023tvh,Basile:2025lwx} and references therein for a partial list). In this paper, we will focus on a particular prediction with dramatic experimental and observational consequences—\emph{an effective reduction in the strength of gravity at small scales.} 

In \cite{Danielsson:2023alz}, it was argued that the dark bubble hypothesis leads to a natural hierarchy involving the AdS-scale $L$, the string scale $l_s$, the 4D Planck scale $l_4$, and the 5D Planck scale $l_5$. This is based on dimensional reductions from 10D to 5D, as well as the way in which 4D gravity appears on the brane through the junction conditions. In \cite{Danielsson:2023alz}, a connection was also made with the positive cosmological constant induced in 4D such that all scales can be determined by a single large number $N$. Expressed in terms of the 4D Planck length $l_4$, we find the string scale to be of order $l_s \sim N^{1/4} l_4$, the AdS-scale to be of order $L \sim N^{1/2} l_4$, and the cosmological scale to be of order $N l_4$. In addition, what is unique about the dark bubble is that \emph{the 5D Planck length is smaller than the 4D Planck length}, and of order $l_5 \sim N^{-1/6}l_4$.

Fixing $N\sim 10^{60}$ through the measured cosmological constant suggests that the AdS-scale should be of order $10^{-5}$m. If so, one expects there to be modifications of the force of gravity that can be measured using table top experiments. Amusingly, this length scale is logarithmically half way between the 4D Planck scale and the cosmological horizon scale.\footnote{The string scale is predicted to sit somewhere around $10$TeV—logarithmically halfway between microns and the 4D Planck scale. We will use natural units $\hbar=c=1$ throughout this paper.} Intriguingly, present experimental limits on what scales deviations from the expected $1/r^2$ Newtonian inverse square law can set in are of precisely this order of magnitude. See \cite{Murata:2014nra} for a recent comprehensive review.

In this paper, we study a localized lump of matter on top of the dark bubble. We calculate the 5D backreaction and deduce the 4D metric. Some initial steps were taken in this direction in \cite{Banerjee:2020wix}, but in this paper we resolve some ambiguities and present a unique result for a modified gravitational potential. In \cref{sec:matter-on-the-brane} we review earlier results and, in particular, consider the case of point masses and calculate the modified gravitational potential. As we will see, its qualitative behavior is rather simple. What happens is that the 4D gravitational force turns itself off at distances of order $10^{-5}$m, with only a weak 5D remainder left. In \cref{sec:why-no-eft} we try to interpret the results and comment on whether there is any effective field theory (EFT) for the dark bubble. In \cref{sec:cosmology} we show that the same reduction in the strength of gravity also appears in early universe cosmology. When the Hubble scale becomes sufficiently small, the time evolution is completely changed. It turns out that the Hubble scale becomes constant during an extended period, mimicking inflation. This suggests new scenarios for the origin of the universe, which also explains the present matter content of the universe. Finally, in \cref{sec:blackholes} we speculate on what happens to  sufficiently large black holes in the dark bubble model, before we end with some conclusions in \cref{sec:conclusion}.

\section{Matter on the brane}\label{sec:matter-on-the-brane}

The dark bubble model—which realizes \emph{dark} energy in 4D via \emph{bubble} nucleation in 5D—not only makes a small positive cosmological constant natural, it makes it inevitable.
The introduction of matter is less technically straightforward. Let us briefly review the main results so far.

The dark bubble model is based on two key sets of equations. First, the Israel junction conditions, which relate the jump in the extrinsic curvature across the brane to the energy density of the brane. Second, the Gauss-Codazzi equations, which relate the intrinsic and extrinsic curvatures of the brane to the bulk geometry. Together, these equations give rise to the 4D Einstein equations for the induced geometry on the dark bubble.

Let us focus on time-dependent solutions that are homogeneous and isotropic in 4D, i.e., cosmology. We consider the case with a higher dimensional black hole at the center of the dark bubble, with metric
\begin{align}\label{eq:global-ads5}\nn
    ds^2_\pm &= -\left(k_\pm^2z^2+1-2G_5M_\pm/z^2\right)dt^2\\
    &+\frac{dz^2}{k_\pm^2z^2+1-2G_5M_\pm/z²}+z^2 d\Omega^2_3\,,
\end{align}
on the inside ($-$) and on the outside ($+$).\footnote{In $D$-dimensions, $M \coloneqq \frac{8π \mathcal{M}}{(D-2)\, ω_{D-2} }$ in terms of the asymptotic mass $\mathcal{M}$, where $ω_{D-2} \coloneqq \frac{2π^{(D-1)/2}}{Γ \left(\frac{D-1}{2}\right)}$ is the area of a unit $D-2$ sphere. For $D=5$, this gives $M = (4/3π) \mathcal{M}$.}
Parameterizing the location of the brane  in terms of proper time for an observer at rest on the bubble as $z= a(τ)$, the induced metric takes the FLRW form,
\begin{equation}\label{eq:FLRW}
    \mathrm{d}s² = - \mathrm{d}τ² + a(t)² \mathrm{d}\Omega_3²\,.
\end{equation}
The first junction condition demands continuity of the metric across the brane, while the second junction condition can be written as
\begin{equation}
    \sigma =\frac{3}{8\pi G_5a}\left(\beta_--\beta_+\right) ,
\end{equation}
where $\sigma$ is the tension of the brane supporting the dark bubble.
For convenience, we have defined
\begin{equation}\label{eq:beta}
\beta_\pm=\sqrt{\alpha_\pm+\dot{a}^2} ,
\end{equation}
with
\begin{equation}\label{eq:alpha}
\alpha_\pm=k_\pm^2a^2+1-\frac{2G_5 M_\pm}{a^2},
\end{equation}
for a brane located at $z=a(τ)$.
Following \cite{Brown:1988kg} we multiply the junction condition with the average $2\pi^2a^3\left(\beta_-+\beta_+\right)/2$ across the brane and find 
\begin{align}\label{eq: Hbulk}\nn
    0 &=\pi^2 a^3\left(\beta_-+\beta_+\right)\sigma -\frac{3\pi^2 a^2}{8\pi G_5}\left(\beta_-^2-\beta_+^2\right)\\ \nn
    &=\pi^2 a^3\left(\beta_-+\beta_+\right)\sigma -\frac{3\pi^2}{8\pi G_5}\left[(k_-^2-k_+^2)a^4+2G_5M_+\right]\\ \nn
    &=\pi^2 a^3\left(\beta_-+\beta_+\right)\sigma -\frac{3\pi^2(k_-^2-k_+^2)a^4}{8\pi G_5}+\mathcal{M_-}-\mathcal{M_+}\\ 
    & \equiv {\cal H},
\end{align}
which is just energy conservation in the bulk. The first term is the energy of the brane, while the second is the change in vacuum energy and the last term is the change in the mass of the black hole when the brane nucleates.

We now assume $a(τ) \gg 1$ and find the first Friedmann equation in the form
\begin{align}\label{eq:Friedmann} \nn
\frac{\dot{a}^2}{a^2} \approx -\frac{1}{a^2}+\frac{8\pi G_4}{3} &\Bigg[ \frac{3(k_--k_+)}{8\pi G_5}-\sigma \\
&+\frac{3}{8\pi a^4}\left(\frac{M_+}{k_+}-\frac{M_-}{k_-}\right) \Bigg]\,,
\end{align}
where an overdot represents a derivative with respect to proper time $τ$ on the moving brane, and we have identified the 4D Newton's constant through
\begin{equation} \label{eq: G4}
G_4=\frac{2k_- k_+}{k_- - k_+} G_5 ,
\end{equation}
and read off the dark energy from
\begin{equation}\label{eq:rho-lambda}
    \rho_\Lambda = \frac{3(k_--k_+)}{8\pi G_5}-\sigma  .
\end{equation}
The critical tension is given by $\sigma_c \coloneqq 3(k_--k_+)/(8\pi G_5)$, and the tension $σ$ of the brane needs to be {\it less} than this value for a bubble to be able to nucleate. This implies that the cosmological constant must be positive for the dark bubble to exist. In this way, the dark bubble model {\it predicts} the cosmological constant to be positive.

From the 4D point of view, we note the presence of radiation in this particular setup. This dark radiation is purely due to non-trivial contributions from the bulk geometry. There is nothing actually sitting on the brane. It is important to note that the mass parameters $M_\pm$ need not be the same on the two sides of the brane. The total asymptotic mass $M_+$ can be viewed as the sum of the actual mass $M_-$ of the black hole, the mismatch between the cosmological constants on the the two sides of the brane, the tension of the brane, and the kinetic energy of the brane.

Actually, we can construct a different solution by turning off $M_-$ on the inside and replace it with radiation on top of the brane.  We make the replacement
\begin{equation} \label{eq:replacerad}
   -\sigma+\frac{3}{8\pi a^4}\left(\frac{M_+}{k_+}-\frac{M_-}{k_-}\right) \rightarrow -(\sigma+\rho_r)+\frac{3}{8\pi a^4}\frac{M_+}{k_+}\,,
\end{equation}
with $\rho_r=3 M_-/(8\pi a^4 k_-)$.
This demonstrates a general feature of the dark bubble model. Already when we considered the cosmological constant in \cref{eq:rho-lambda}, we saw that {\it adding} energy to the brane {\it reduces} the cosmological constant. As we have seen, this is simply because bubble nucleation becomes easier when the tension of the brane is reduced. When one wants to add ordinary matter or radiation to the model, this is confusing at first. Looking at \cref{eq:Friedmann} we see that the radiation on top of the brane contributes with a {\it negative} sign to the effective 4D energy density. We immediately recognize the solution to the problem. The last term in \cref{eq:replacerad} represents the gravitational bulk response to the presence of the matter on the brane. It contributes with a positive term that reverses the sign of the direct contribution and gives rise to a net positive energy density. Later in the paper, we will see how this works out in detail.

The above construction assumed the presence of radiation, i.e. massless particles, on the brane. What if we want to consider massive particles? This turns out to be more tricky. The characteristic $1/a^4$ behavior of the density as the bubble expands, is due to the bulk AdS-black hole metric. With massive particles, we expect the density to behave as $1/a^3$. In \cite{Banerjee:2018qey} (see also \cite{Banerjee:2019fzz,Banerjee:2020wov}) it was suggested to use stretched strings. A cloud of such strings leads to $1/a^3$, and one can also show that single strings pull on the brane in such a way that they effectively show up as particles with mass $L\tau$, where $L$ is the AdS scale and $\tau$ is the tension of the string.

In the following, we will demonstrate that this is not the only possibility. In fact, localized massive particles on top of the brane, with the asymptotic metric tuned as above, will source 4D gravity in the expected way. We will consider particles with mass $M$ such that $G_4 M \ll L$. That is, their Schwarzschild radius is smaller than the AdS-scale. As we will see, gravity will be cutoff in such a way that the linear approximation to Einstein's equations will always be sufficient. In particular, we will find that no horizon will ever develop in 4D for such small masses.

Let us begin with a review of the general case.

\subsection{Solving the junction conditions}

Let us embed a spherical brane with tension $\sigma$ in the 5D background in \cref{eq:global-ads5} such that 
\begin{equation}\label{eq:fpm}
    z(r) = z_0(1 + f_\pm(r)) .
\end{equation}
It will be sufficient to work at linear order in $f_\pm (r)$. Inside the brane bubble we have $k=k_-$, while on the outside we have $k=k_+ < k_-$. According to Israel's junction conditions,  the jump in the extrinsic curvature across the brane will be related to its energy momentum tensor. We define
\begin{equation}\label{eq:Sab-def}
S_{ab} \coloneqq K_{ab}-K h_{ab},
\end{equation}
and write
\begin{equation}
    \sigma h_{ab}+T_{ab}^{(p)}= -\frac{1}{8\pi G_5} \left(S^{(-)}_{ab} -S^{(+)}_{ab}  \right).
\end{equation}
We have separated out the tension of the brane, so that $T_{ab}^{(p)}$ is the extra energy density on top of the brane. This is {\it not} the same as the energy density measured by a 4D observer. According to the Gauss-Codazzi equation
\begin{equation}
    R_{ab}^{
    (4)}+ K_a^c K_{cb} -K K_{ab} ={\cal I}_{ab},
\end{equation}
where
\begin{equation}
    {\cal I}_{bd} \coloneqq g^{ac} R^{(5)}_{\alpha \beta \gamma \delta}e_a^{\alpha}e_b^{\beta}e_c^{\gamma}e_d^{\delta}\,,
\end{equation}
and $e^α_a \coloneqq ∂x^α/∂y^a$ are tangent to the brane, where $x^α$, and $y^a$ are coordinates in the 5D bulk and on the 4D brane respectively.
With this, the junction conditions can be identified with the 4D Einstein equation $G_{ab} = 8 \pi G_4 T^{(a)}_{ab}$. Here, $T^{(a)}_{ab}$ is the energy momentum tensor that is actually measured by the 4D observer.\footnote{The superscripts $^{(a)}$ and $^{(p)}$ on $T_{ab}$ stand for ``active" and ``passive" stress-tensors. We will explain this in \cref{sec:why-no-eft}.} For simplicity, we will consider a critical brane with tension $σ_c$, defined below \cref{eq:rho-lambda}, so that the 4D cosmological constant vanishes. If the tension is slightly less than the critical value (making it possible for a dark bubble to nucleate) the cosmological constant will be small and positive.
Assuming a small matter density on the brane, we then find \cite{Banerjee:2019fzz, Banerjee:2022ree}
\begin{equation} \label{eq.Einstein4D}
    T^{(a)}_{ab} = -T^{(p)}_{ab} + \frac{{\cal I}^{(+)}_{ab}}{k_+}-\frac{{\cal I}^{(-)}_{ab}}{k_-}-\frac{1}{2}\left( \frac{{\cal I}^{(+)}}{k_+}-\frac{{\cal I}^{(-)}}{k_-}
    \right) h_{ab}\,.
\end{equation}
What happens is that a mass on the brane, naively contributes with a {\it negative} sign to the net 4D energy density. This is the first term on the right hand side. The terms involving ${\cal I}_{ab}$ give the back reaction from bending of the brane in the bulk \cite{Garriga:1999yh,Giddings:2000mu}. Summing the two contributions leads to a {\it positive} energy density. We have already seen how this works in the case of radiation.
The matter on the brane pushes the brane down, while the gravitational back reaction from the bulk pulls the brane up \cite{Banerjee:2020wix}.

What about a point mass? As already mentioned, the problem was solved in \cite{Banerjee:2018qey}  not by a point mass on the brane, but by a string pulling on the brane. As observed in \cite{Banerjee:2019fzz}, if the tension of the string is order the string scale, the mass is of order Planck mass. From the point of view of (\ref{eq.Einstein4D}), the pulling of the string is accounted for by the tensor ${\cal I}_{ab}$, capturing the bulk curvature.

\begin{figure}
    \centering
    \includegraphics[width=\linewidth]{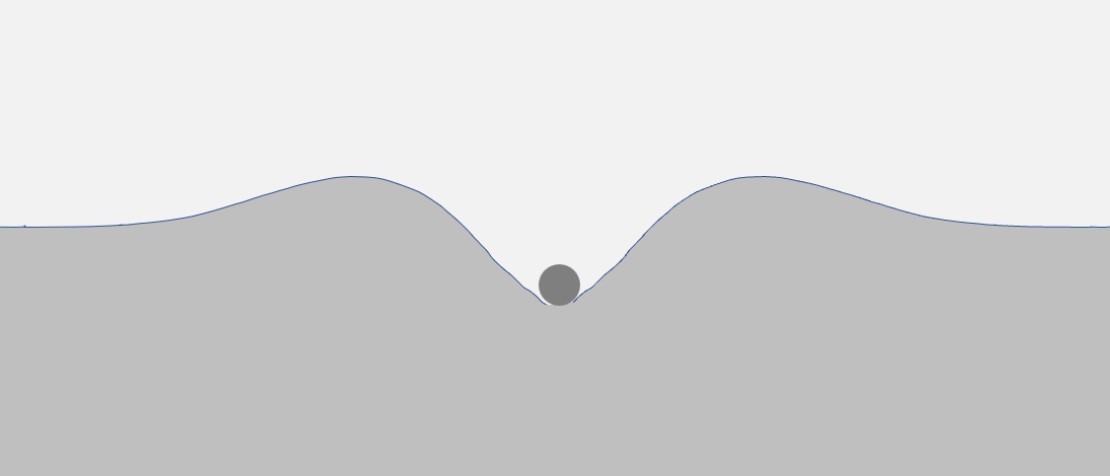}
    \caption{A small mass resting on the dark bubble.}
    \label{fig:matteronbrane}
\end{figure}

What happens if we add a point mass directly on the brane? It is no longer obvious how the pulling up actually is realized, but there is an intuitive argument for why the effect has to be there. Imagine a pressurized balloon representing the dark bubble. Put a small stone on top of it. The surface will be pushed down at the location of the stone, but if the volume of the inside is supposed to be constant, it will press itself up in a region further out. See \cref{fig:matteronbrane}.

In case of a mass on the dark bubble, this simple argument suggests that a localized mass on the brane can create its own uplift. We will now demonstrate that this is in fact what happens, provided that the right boundary conditions are chosen.

\subsection{General solution for 5D metric}

Consider the following asymptotically AdS bulk metric with a spherically symmetric ansatz in 4D spacetime, where $h_i$ are the perturbation on top of AdS
\begin{align}\nn
    ds_5^2 &= g_{\mu \nu}dx^\mu dx^\nu\\ \nn
    &=\left(-k^2 z^2+ h_t(r,z) \right) dt^2 + \left(k^2 z^2 + h_r(r,z) \right)dr^2\\
    &+\left(k^2 z^2 + h_a(r,z) \right)r^2\left(d \theta^2  +  \sin^2 \theta d\phi^2 \right) +\frac{dz^2}{k^2 z^2},
\end{align}
and we choose a gauge in which the perturbation is traceless
\begin{equation}\label{eq:trace}
h_r(r,z)+2h_a(r,z)-h_t(r,z)=0.
\end{equation}

For $h_i=0$, this is the AdS$_5$ metric in the Poincaré patch with $z→∞$ being the boundary of AdS, and $z→0$ the horizon.
At linear order in $h_i$, Einstein's equation determining $h_t$ is given by
\begin{multline}\label{eq:einstein-ht}
    ∂_{rr} h_t(r,z) + \frac{2}{r}∂_r h_t(r,z) - 4k^4 z^2 h_t(r,z)\\ + k^4 z^3 ∂_z h_t(r,z)+ k^4 z^4 ∂_{zz} h_t(r,z) = 0.
\end{multline}
Defining a Fourier sine  transform of $h_t(r,z)$ as $h(p,z)$ using the following,
\begin{equation}\label{eq:ht-hp}
    h_t (r,z) = \frac{1}{r} \sqrt{\frac{2}{\pi}}\int_0^\infty dp\ h(p,z) \sin (p r),
\end{equation}
\cref{eq:einstein-ht} becomes
\begin{equation}
    z^2 ∂_{zz} h(p,z) + z ∂_z h(p,z) - \left( 4 +\frac{p^2}{k^4 z^2} \right) h(p,z) = 0.
\end{equation}
In terms of the variable $p/(k^2 z)$, this is simply the Bessel equation
which has solutions in terms of Bessel functions $K_2$ and $I_2$:
\begin{equation}\label{eq:hp-solution}
    h (p,z)= A(p) K_2 \left(\frac{p}{k^2 z}\right) + B(p)I_2\left(\frac{p}{k^2 z}\right).
\end{equation}
Note that while $h(r,z)$ is dimensionless, the Fourier transforms $h(p,z)$ as well as $A(p)$ and $B(p)$ carry dimension length$^2$.
The other independent metric function $h_r(r,z)$ is determined in terms of $h_t(r,z)$ by 
\begin{equation}
    r ∂_r h_r(r,z) + 3 h_r(r,z) - h_t(r,z) = 0.
\end{equation}
Using \cref{eq:ht-hp}, this gives
\begin{equation}
    h_r (r,z) = \frac{1}{r^3}\sqrt{\frac{2}{\pi}} \int_0^\infty dp\, h(p,z) \frac{\sin (p r)-pr \cos (pr)}{p^2}\,.
\end{equation}
So the metric is completely determined by a single function $h(p,z)$ in \cref{eq:hp-solution}.

In holographic language, the second term in \cref{eq:hp-solution} is the normalizable solution, while the first term is the non-normalizable one. For us, the most important property of the non-normalizable solution is that it is finite as $z → 0$ (which corresponds to the center of the bubble), while it blows up as $z → \infty$. The normalizable solutions have the exact opposite properties. In particular, the black hole in AdS discussed previously is a normalizable solution. In the case of the homogeneous, cosmological and time dependent case with radiation described earlier, there are only normalizable solutions at play. Demanding the interior to be non-singular the interior contains only the $K_2$ mode.

With a static localized solution, we will again have normalizable modes on the outside but not on the inside. In addition, we must now take into account non-normalizable solutions on the inside as well as on the outside. That is, we will have non-zero $A_-(p)$, $A_+(p)$ and $B_+(p)$.
\begin{figure}
    \centering
    \includegraphics[width=0.9\linewidth]{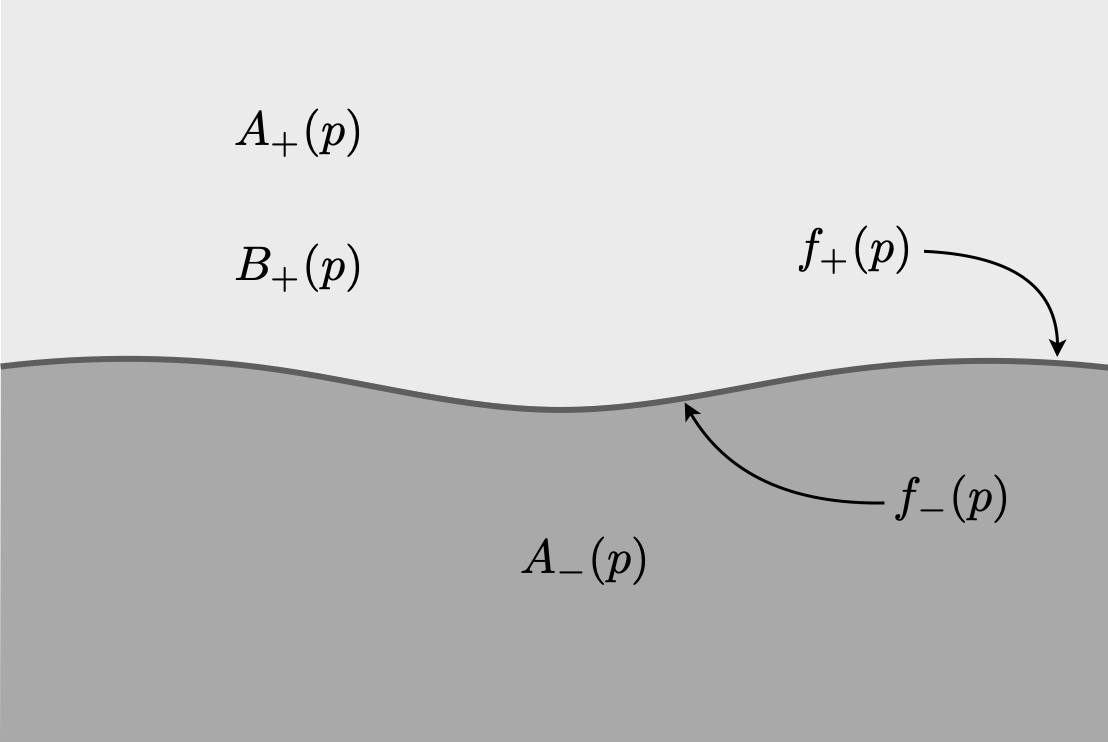}
    \caption{Spacetime outside the bubble (top of the figure, in light gray) contains both the normalizable $I_2$ (decaying towards the boundary) and the non-normalizable $K_2$ (growing towards the boundary) modes, corresponding to the functions $B_+(p), A_+(p)$ respectively. Regularity at the center of the bubble requires that the inside (bottom half of the figure, in dark gray), contains only the non-normalizable $K_2$, corresponding to $A_-(p)$. The functions $f_\pm(p)$ determine how the brane is embedded into the bulk.}
    \label{fig:AplusBplus}
\end{figure}
The functions that we need to determine are—$A_+(p)$, $B_+(p)$, and $A_-(p)$ in the metric, together with $f_-(p)$ and $f_+(p)=f_-(p)+df(p)$ from the embedding defined in \cref{eq:fpm}—five functions in total, as shown in \cref{fig:AplusBplus}. To our help we have the first junction condition, one constraint, and the second junction condition, which will give two constraints from the energy density and the pressure on the brane. Across the brane we match the radial coordinate through
\begin{equation}
    \rho= k_- z_0 r_-=k_+ z_0 r_+,
\end{equation}
where $\rho$ is the proper 4D radius on the brane. Similarly, we note that the proper 4D momentum is given by
\begin{equation}
    q=\frac{p_-}{k_- z_0}=\frac{p_+}{k_+ z_0}.
\end{equation}
We express all momentum dependent functions in terms of $q$ already from the start.\footnote{By a slight abuse of notation we write $A_+(p)=A_+(q)$ etc. We also make a rescaling so that the Fourier transforms use $q$ on both sides.} The energy component of the first junction condition then gives
\begin{align} \label{eq:firstjunction}\nn
    B_+(q) &=\frac{A_-(q)K_2(q/\km)-A_+(q)K_2(q/\kp)+2df(q)}{I_2(q/\kp)}\\
    &\equiv \frac{A_-(q)K_2^--A_+(q)K_2^+ +2df(q)}{I_2}\,,
\end{align}
where we have defined $K_i^\pm \coloneqq K_i(q/k_\pm)$, and  $I_i \coloneqq I_i(q/k_+)$ for brevity, and will continue using this abbreviated notation throughout.

Next, we turn to the second junction condition. It is useful to introduce the following basis
\begin{align}\label{eq:G-matrices}\nn
G_1&=   \frac{\sin q \rho}{\rho}\begin{pmatrix}
 1 & 0 & 0 & 0\\
0 & 0 & 0 & 0 \\
0 & 0 & 1 & 0 \\
0 & 0 & 0 & 1
\end{pmatrix},
 G_2 =   \frac{\sin q \rho}{\rho}\begin{pmatrix}
 1 & 0 & 0 & 0\\
0 & 0 & 0 & 0 \\
0 & 0 & -\frac{1}{2} & 0 \\
0 & 0 & 0 & -\frac{1}{2}
\end{pmatrix}\,,\\
 G_3 &=   \frac{q \rho\cos p\rho -\sin p\rho}{q^2 \rho^3}\begin{pmatrix}
 0 & 0 & 0 & 0\\
0 & 1 & 0 & 0 \\
0 & 0 & -\frac{1}{2}  & 0 \\
0 & 0 & 0 &  -\frac{1}{2}
\end{pmatrix}\,.
\end{align}
The jump in the trace-reversed extrinsic curvatures defined in \cref{eq:Sab-def} is given by
\begin{align}\label{eq:Sab-diff}\nn
    &S^{(+)a}_b -S^{(-)a}_b = 3(k_- - k_+) \times \mathbbm{1} \\ \nn
    &-\sqrt{\frac{2}{\pi}}\int_0^\infty dq\frac{F(q)}{k_-k_+} \left(q² G_1 - 2q² G_3 \right) +\sqrt{\frac{2}{\pi}}\int_0^\infty \frac{qdq}{2} \times \\
    &\times \left(B_+(q)I_1 +A _-(q)K_1^- -A_+(q)K_1^+\right)(G_2 + G_3)\,,
\end{align}
and by the second junction condition, the stress-tensor on the brane is $(8πG_5)^{-1}$ times the above.
Here we have defined \footnote{This is a generalization from \cite{Padilla:2004mc}, where it was assumed that $k_-f_-(q)=k_+f_+(q)$.}
\begin{equation}\label{eq:Fofq}
    F(q)=k_- f_+(q)-k_+f_-=(k_--k_+)f_-(q)+k_-df(q).
\end{equation}
These equations do not fully determine the solutions and are fairly complicated. The most difficult step, when the embeddings are different on the two sides of the brane, is to solve the first junction condition. Luckily, this is not necessary. We will soon see how the bending of the brane can be completely removed through changes of coordinates in the bulk. This will still not fix all unknown functions, but in the next section we will explain how to obtain the one remaining condition.

\subsection{Dark bubble holography}

To find the physically relevant solution we will need one more condition. This will be imposed through holographic renormalization using a non-dynamical cutoff. We will call the theory at the cutoff, the ``hologram". Note that the hologram is not a dynamical brane, but just an arbitrary cutoff at which we specify boundary conditions on the bulk fields. What we will find is a generalization of the ordinary AdS/CFT correspondence. There, we impose Dirichlet boundary conditions at the boundary at infinity, which are encoded into a CFT. The non-normalizable modes parametrize different theories, while the normalizable ones correspond to different expectation values within the same theory. To the CFT induced on the cutoff brane through the extrinsic curvature in York term, we also need to add counter terms. Usually, these terms are designed to cancel all terms except those belonging to the CFT. In the case of the dark bubble the non-normalizable modes will be dynamical and we need to impose mixed boundary conditions rather than pure Dirichlet, \cite{Banerjee:2023uto}. As we will go through in detail in the next section, the subtraction on the cutoff brane is adjusted to leave a nontrivial Hilbert term in the action. Following \cite{Compere:2008us, Banerjee:2012dw, Banerjee:2023uto}, we impose
\begin{equation}
T_{ab}^{(\textrm{\scshape cft})}+T_{ab}^{(\textrm{ct})}=T_{ab}^{(\textrm{reg})},
\end{equation}
where the counter terms are given by
\begin{equation}
    T_{a}^{b(\textrm{ct})}= \kappa_1 G_a^b + \kappa_2 \delta_a^b.
\end{equation}{}
If we were to choose $T_{ab}^{(\textrm{reg})}=0$, we would find Neumann boundary conditions, while a non-zero value leads to mixed conditions.

To obtain the needed conditions, \emph{we will impose that the hologram is a perfect representation of the theory on the brane}. Mathematically, this is not a necessary condition. In principle we could have all kinds of structures in the bulk that would make the hologram very different with a complicated dependence on the location of the cutoff brane, $z=z_c$. What we will do is to assume that the bulk is as empty and simple as possible. Amusingly, there is a parallel with the notion of non-renormalizability of gravity. The statement that gravity cannot be renormalized does not imply that the theory is inconsistent. It just means that it lacks predictability. At every step of perturbation theory on the way to higher energies, we'd need to introduce more unknown coefficients. We consider this to be a parallel with the dark bubble and the arbitrariness of the bulk. Luckily, there is a simple possibility that suggests itself, which we will focus on.

Before considering localized matter sources on the brane, let us consider the case of homogeneous and isotropic radiation that we reviewed earlier. It is natural to impose  a time independent cutoff in the bulk at a constant value $z_c \gg z_0$. If $z_0=z_0(t)$, with the expansion of the universe taken into account, such a cutoff would evolve down in energy with time from the 4D point of view. On the other hand, we could also let $z_c=z_c(t)$ evolve in time in 5D, so that $z_c(t)/z_0(t)$ remains time independent. This would correspond to a time independent cutoff in 4D. We will discuss both kinds of cutoffs below.

\subsubsection{Time independent 4D-cutoff}

The procedure follows that of \cite{Banerjee:2023uto}. We start by recalling the junction conditions on the dark bubble, which give rise to the 4D Einstein equation,
\begin{align}\nn
  &\sigma+\rho_r = \frac{3}{8\pi G_5 z_0} \sqrt{k_-^2 z_0^2+\dot{z}_0^2+1-2G_5M_-/z_0^2}\\ \nn
  &\qquad \quad- \frac{3}{8\pi G_5 z_0} \sqrt{k_+^2 z_0^2+\dot{z}_0^2+1-2G_5M_+/z_0^2} \\ \nn
  &\approx \frac{3k_-}{8\pi G_5}-\frac{3k_+}{8\pi G_5}+\frac{3}{16\pi k_- G_5}\left(\frac{\dot{z}_0^2}{ z_0^2}+\frac{1}{z_0^2} -\frac{2G_5M_-}{z_0^4}\right)\\
  &\qquad \qquad -\frac{3}{16\pi k_+ G_5}\left(\frac{\dot{z}_0^2}{ z_0^2}+\frac{1}{z_0^2} -\frac{2G_5M_+}{z_0^4}\right) .
\end{align}
If we choose $M_-=0$, this becomes
\begin{align}\label{eq:tmunu-darkbubble}\nn
 &\frac{3}{16\pi G_5}\left(\frac{1}{k_+}-\frac{1}{k_-}\right)\left(\frac{\dot{z}_0^2}{ z_0^2}+\frac{1}{z_0^2} \right)\\
 &\qquad \qquad\approx \frac{3k_-}{8\pi G_5}-\frac{3k_+}{8\pi G_5}-\sigma+\frac{3M_+}{8\pi k_+  z_0^4}-\rho_r .
\end{align}
The goal is now to reproduce this equation using $T_0^{0(\textrm{reg})}$ for some choice of the coefficients $\kappa_{1,2}$. On the dark bubble we balance the extrinsic curvature of the interior against the extrinsic curvature of the exterior. In the hologram there is no junction condition, just an arbitrary cutoff on which we try to define an effective theory. The counterterms effectively serve the same role as the extrinsic curvature of the interior does for the dark bubble. We have
\begin{align}\nn
     \epsilon &\coloneqq T_0^{0(\textrm{reg})} 
     = \kappa_2 - 3\kappa_1 \left(\frac{\dot{z}_c^2}{ z_c^2}+\frac{1}{z_c^2}\right)\\ \nn
     &\qquad \qquad \quad-\frac{3}{8\pi G_5 z_c}\sqrt{k_+^2 z_c^2+\dot{z}_c^2+1-\frac{2G_5M_+}{z_c^2}} \\ \nn
    &\approx \kappa_2 -\frac{3k_+}{8\pi G_5}- 3\kappa_1 \left(\frac{\dot{z}_c^2}{ z_c^2}+\frac{1}{z_c^2}\right)\\ 
    &\quad -\frac{3}{16\pi k_+ G_5}\left(\frac{\dot{z}_c^2}{ z_c^2}+\frac{1}{z_c^2} -\frac{2G_5M_+}{z_c^4}  \right) + \mathcal{O} \left( \frac{1}{k_+³ z_c³}\right)\,,
\end{align}
which can be written as
\begin{multline} \label{eq: rad-reg}
    \left(\frac{3}{16\pi k_+ G_5}+3\kappa_1\right)\left(\frac{\dot{z}_c^2}{ z_c^2}+\frac{1}{z_c^2} \right)\\
    \approx \kappa_2-\frac{3k_+}{8\pi G_5}+\frac{3M_+}{8\pi k_+  z_c^4}-\epsilon\,.
\end{multline}
Note that the proper time is calculated using $z_c$ in the hologram, and $z_0$ on the dark bubble. With $k_+ z_c dt= d\tau_c$ and $k_+ z_0 dt = d\tau_0$, recalling that the ratio $z_c/z_0$ is time independent for a time independent 4D cutoff, we obtain $d z_c/d\tau_c=(z_c/z_0)d z_0/d\tau_c=d z_0/d\tau_0$. Hence it follows that $\dot{z}_c^2/z_c^2=(z_0^2/z_c^2) H^2$. For the energy density in the hologram, \cref{eq: rad-reg}, to match that on the dark bubble \eqref{eq:tmunu-darkbubble} up to scaling by $(z_c/z_0)^4$, we need to pick
\begin{align}\label{eq:kappa1}\nn
    \kappa_1 &= -\frac{1}{16\pi G_5 k_+}+\frac{\Delta \kappa_1 \Delta k z_0^2}{16 \pi G_5 k_+ k_- z_c^2}\\
    &=-\frac{1}{16 \pi G_5 k_+}+\frac{\Delta \kappa_1 z_0^2}{8 \pi G_4 z_c^2}\,,\\ \label{eq:kappa2}
    \kappa_2 &= \frac{3k_+}{8\pi G_5}+\frac{\rho_\Lambda  z_0^4}{z_c^4}\,,
\end{align}
where a non-zero value of $\Delta \kappa_1$ takes us away from the critical value $κ_1 = -(16\pi G_5 k_+)^{-1}$, where gravity would have been completely canceled. To obtain this we have used (\ref{eq: G4}) and (\ref{eq:rho-lambda}).

We need $\Delta \kappa_1=1$ to reproduce the expected 4D gravity. This implies that Newton's constant in the hologram is a factor $z_c^2/z_0^2$ larger than on the dark bubble. That is, the Planck length is larger by a factor $z_c/z_0$. We also note that the regulated energy momentum tensor $T_0^{0(\textrm{reg})}$ is given by the energy density on the brane, so that $\epsilon=\rho_r$. This is {\it not} what a 4D observer would measure. This, instead, is given by $\rho_4 \sim M_+/(k_+z^4)-\rho_r$, where we still have not fixed the value of $M_+$. We will come back to this after we have discussed the localized case. 
Let us now briefly comment on time dependent cutoffs.

\subsubsection{Time dependent 4D-cutoff}

Let us introduce a time independent cutoff in 5D at a constant $z=z_c \gg z_0(t)$. The value of $z_c$ is arbitrary as long as it is large. As the universe expands, we might at some point need to increase it. Beyond this, the actual value of the cutoff should not affect the 4D physics. At the expanding non-dynamical cutoff, there will be a holographic representation of the interior 5D-theory, which will mirror the physics on the dark bubble. From a 5D point of view, this may seem to be the most natural kind of cutoff. However, the price to pay is that \emph{the cut-off becomes time dependent in 4D}.

The metric of the hologram is not identical to the one on the dark bubble since the dark bubble is non-trivially embedded through $z_0(t)$. To represent this degree of freedom, we need to take into account a conformal factor $z_c/z_0(t)$ in the hologram. The result is a theory where the 4D Planck length $l_4$ depends on $z_c$. That is, when we increase $z_c$ and move further up in the throat toward the UV, $l_4$ becomes {\it larger}. On the other hand, if we keep $z_c$ fixed and let $z_0(t)$ increase with time, as the universe expands, $l_4$ becomes {\it smaller}. This makes perfect sense. The hologram is an alternative way to represent our universe, where the size remains fixed but the Planck length decreases with time.

To be more precise, the time dependent scale factor in the metric on the dark bubble is induced from the embedding $z_0(\eta)$, where we use conformal time $\eta$. The Hilbert term corresponding to \cref{eq:FLRW} in conformal time is simply given by 
\begin{equation} \label{eq: R dark bubble}
    \int \sqrt{-g}\,R=\int \sqrt{-g}\,\frac{6(a+a'')}{a^3},
\end{equation}
where the scale factor is given by $a(\eta)=z_0(\eta)$. On the hologram, the embedding coordinate $z_c$ is time independent, and the time dependence instead comes from a scalar $\phi (\eta )$ describing a time dependent 4D Planck scale. According to our prescription, the action should be numerically the same, and we require it to decompose as follows.
\begin{equation} \label{eq: R hologram}
    \int \sqrt{-g}\,R = \int \sqrt{-g}\,ϕ²\Bigg[\phi^{-1} \bar{R}+\frac{3\left( 2\phi \phi''- \phi '^2\right)}{2\phi^3}\Bigg], 
\end{equation}
where $\bar{R}=6/z_c²$
is the time independent curvature of the hologram. Up to a total derivative, the theory on the hologram is formally a Brans-Dicke theory at the special value $\omega =-3/2$.
At this value the scalar field $\phi$ does not represent any extra degree of freedom. The action is obtained simply through a conformal field redefinition. We verify that
\begin{equation}
\phi(\eta)=\frac{a(\eta)^2}{z_c^2},
\end{equation}
maps (\ref{eq: R hologram}) onto (\ref{eq: R dark bubble}), taking the rescaling of proper time into account.
In this way, we obtain a conformally rescaled theory on the hologram that perfectly reproduces the physics on the dark bubble. The actual 4D physics is the same as in the previous section. 

\begin{figure}%
    \centering
    \includegraphics[width=\linewidth]{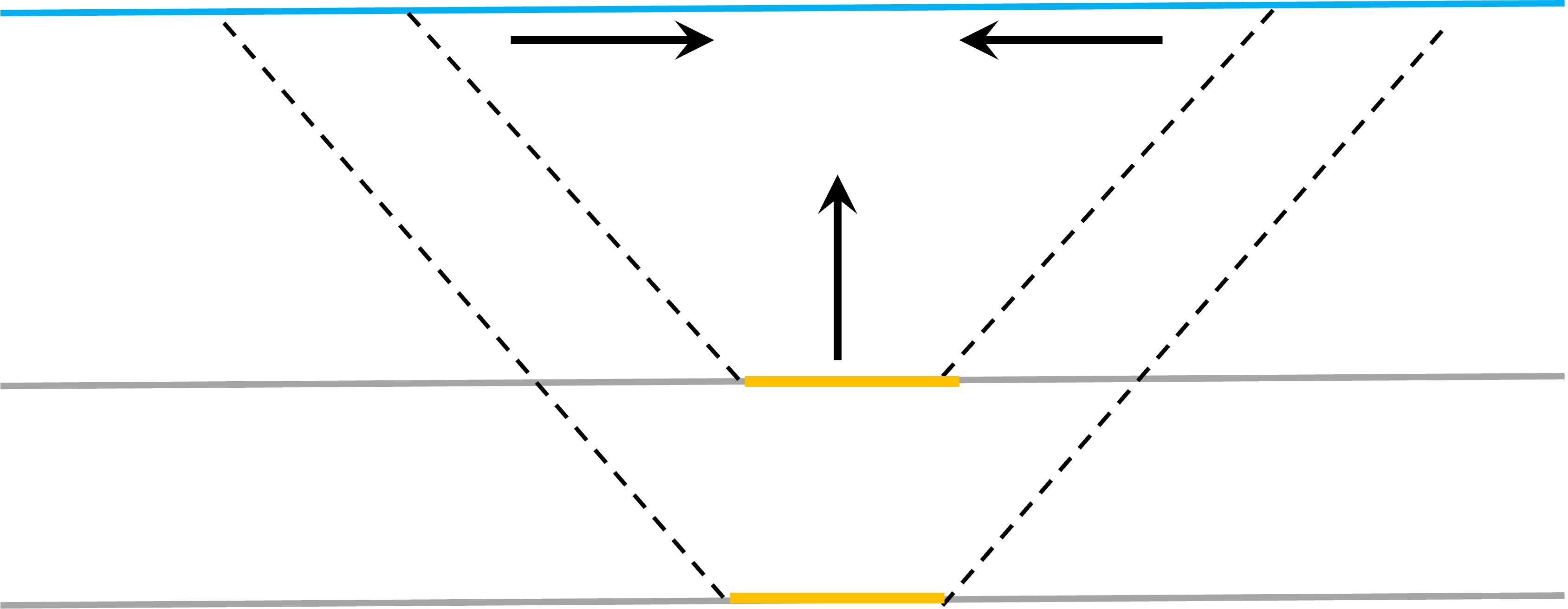}
    \caption{The scale where gravity is turned off is constant on the dark bubble, but its projection in the hologram is shrinking. Thus, in the hologram gravity is turning on starting in the IR. }
    \label{fig:holgrav}
\end{figure}

Amusingly, it is the very presence of the dark bubble that is the reason that there is any gravity at all in the hologram. To understand this, let us start with a holographic CFT without gravity. We assume that the bulk is unstable against the formation of an expanding dark bubble signaling a phase transition. If the cutoff at asymptotic infinity is held at a constant radius, such a phase transition initially affects the boundary hologram at large scales and, as time evolves, on smaller and smaller scales.  Physically, this is interpreted as gravity first being turned on at large length scales and then working itself down to smaller length scales into the UV. As we have discussed, the hologram placed at a constant radius depicts an expanding universe as a static universe with a shrinking Planck length. 

\subsection{Massless sources}

Let us now come back to the case with localized matter on the brane. We still consider traceless matter where $F(q)=0$, so that we can simply choose $f_-(q)=f_+(q)=0$. The time-time component of the stress tensor in \cref{eq:Sab-diff}, then reduces to 
\begin{multline} \label{eq:secondjunct}
   \frac{1}{8\pi G_5 \rho} \sqrt{\frac{2}{\pi}}\int_0^\infty dq\, \sin(q ρ) \times \\
   \times \frac{q}{2} \left(B_+(q)I_1+A_-(q)K_1^--A_+(q)K_1^+\right)=M(\rho) ,
\end{multline}
where $M(\rho)$ is the energy density on the dark bubble.
We conclude that 
\begin{equation}
    \frac{q}{2}\left(B_+(q)I_1+A_-(q)K_1^--A_+(q)K_1^+\right)=8 \pi G_5 M(q) ,
\end{equation}
where $M(q)$ is the Fourier sine  transformed energy density with dimension length$^{-2}$, recalling that $A$ and $B$ have dimensions length$^2$.

Together with the first junction condition, i.e. \cref{eq:firstjunction} with $df(p)=0$, we have two equations for three unknown functions: $A_\pm(q), B_+(q)$. For convenience, we will use expressions with one index up and one down. On the cutoff brane we then have \cite{deHaro:2000vlm}
\begin{equation}
    T_b^{(\textrm{\scshape cft}) a}=\frac{1}{8\pi G_5} S_b^a = a_0 + \frac{a_2}{z_c^2}+\frac{a_4}{z_c^4} + ... 
\end{equation}
It is the last term that yields the effective energy momentum tensor of the CFT, while the first two need to be canceled by the counterterms $T_b^{(\textrm{ct}) a}$. Expanding the extrinsic curvature as well as the counterterms, we find
\begin{align}\nn
     T_0^{(\textrm{\scshape cft}) 0} &=\frac{1}{8\pi G_5}\Bigg[ -3k_+ + \sqrt{\frac{2}{\pi}}\int_0^\infty dp \times \\ \nn
     &\times \frac{p(B_+(p)I_1(\frac{p}{k_+^2z_c})-A_+(p)K_1(\frac{p}{k_+^2z_c}))}{2k_+^3r z_c^3} \sin (p r)\Bigg] \\ \nn
     \xrightarrow{z_c k_+ \gg 1}&\frac{1}{8\pi G_5} \Bigg[-3k_+ +\sqrt{\frac{2}{\pi}}\int_0^\infty dp \times \\ \nn
     &\hspace{-2em} \times \frac{1}{4k_+^3 r z_c²} \left(\frac{p^2}{k_+^2z_c^2}B_+(p)-2k_+^2 A_+(p)\right)\sin (p r)\Bigg]+...\nonumber \\
     T_0^{(\textrm{ct}) 0} &= - \kappa_1 \sqrt{\frac{2}{\pi}}\int_0^\infty dp \times \\ \nn
     &\hspace{-2em}\times \frac{p^2(B_+(p)I_2(\frac{p}{k_+^2z_c})+A_+(p)K_2(\frac{p}{k_+^2z_c}))}{2k_+^4r z_c^4} \sin (p r) + \kappa_2 \nonumber \\
     \xrightarrow{z_c k_+ \gg 1}& -\kappa_1 \Bigg[ \sqrt{\frac{2}{\pi}}\int_0^\infty dp \times \\ \nn
     &\hspace{-2em}\left(\frac{p^4}{16 k_+^8 r z_c^6}B_+(p)+\frac{1}{r z_c²}A_+(p)\right)\sin (p r)+...\Bigg]+ \kappa_2 .
\end{align}
As we will justify in a  moment, we only need to keep the leading terms in $1/z_c$ for the $A_+(p)$ and $B_+(p)$ terms, when determining the coefficients $\kappa_{1,2}$ of the counterterms. For convenience, we use $p \equiv p_+$ instead of $q$, since we evaluate on the outside, away from the dark bubble. In summary, we find
\begin{align} \label{eq: treg}
    T_0^{(\textrm{reg}) 0} &= -\frac{3k_+}{8\pi G_5}+ \kappa_2 \nonumber \\ 
    &\hspace{-2em}+  \sqrt{\frac{2}{\pi}}\int_0^\infty dp  \Bigg[ \left(\frac{p^2}{ 32\pi k_+^5 G_5 z_c^2}-\frac{\kappa_1 p^4}{16 k_+^8 z_c^4}\right)\frac{B_+(p)}{r}\\ \nn
    &\hspace{-2em}-\left(\frac{1}{16\pi k_+ G_5 }+\kappa_1\right)\frac{A_+(p)}{r} \Bigg]\sin (p r)+...
\end{align}
This should be compared with (\ref{eq: rad-reg}) and we again need to choose $\kappa_{1,2}$ such that we obtain a scaled version of the theory on the dark bubble.\footnote{For simplicity, we ignore the 4D cosmological constant in this calculation. For realistic scenarios, the length scale of the 4D cosmological constant  is much larger than the microscopic scales where modifications of gravity set in.} We immediately see that the coefficient $\kappa_2$ must be chosen as in \cref{eq:kappa2} to yield the correct cosmological constant. We again note that there is a critical value of $\kappa_1$ at $-(16\pi G_5 k_+)^{-1}$ such that the non-normalizable term $A_+(p)$ is completely canceled in $T_0^{(\textrm{reg}) 0}$. This value would imply that gravity is completely turned off.

To retain gravity, we must balance the non-normalizable terms against the normalizable ones, involving the $B(p)/z_c^2$ term of $T_0^{(\textrm{\scshape cft}) 0}$. If we tune $\kappa_1$ away from the critical value and set it to
\begin{equation}\label{eq:kappa1-2}
    \kappa_1=-\frac{1}{16\pi G_5 k_+} +\frac{\Delta \kappa_1 z_0^2}{8\pi G_4 z_c^2},
\end{equation}
we find the net result
\begin{equation}\label{eq:ABsign}
     \frac{p^2}{32 \pi k_+^5 G_5 z_c^2} B_+(p)-\frac{\Delta \kappa_1 z_0^2}{8\pi G_4 z_c^2 } A_+ (p)=\frac{ z_0^2}{8\pi G_4 z_c^2 } A_+ (p).
\end{equation}
Here we have imposed that the addition of the counterterm should leave us with 4D gravity in the hologram, where $G_4$ is enhanced through $z_c^2/z_0^2$. As we saw earlier, this is needed to compensate for the $1/z_c^4$ decrease in the energy density such that the dimensionless combination $\epsilon G_4^2$, with $\epsilon$ as the energy density, is invariant. That is, physics is invariant up to scaling. Note that the subleading term $\sim 1/z_c^2$ in $A_+(p)$, is multiplied with an extra factor $1/z_c^2$ from \cref{eq:kappa1-2} in \cref{eq: treg}. This justifies why we could drop it when finding $\kappa_1$.

This is the net Einstein tensor, while $T_0^{(\textrm{reg})0}$ becomes the actual matter density on the brane.\footnote{Note that in the limit $z_c\rightarrow\infty$ we have $A_+(p)\rightarrow A_-(p)$, so that we find a match with the theory on the dark bubble.} Note that with $\Delta \kappa_1=1$, the normalizable term with $B_+(p)$ inverts the sign of the Einstein term exactly in the same way as in the Friedmann equation. The remaining, subleading, normalizable term with $B_+(p)$ corresponds to effective energy-momentum on the dark bubble. Changing the non-normalizable terms corresponds to changing the boundary theory.

To summarize, we get a relation between $B_+(p)$ and $A_+(p)$ such that 
\begin{align}\label{eq:Bcond}\nn
B_+(p)&=\frac{4k_+^5 z_0^2 G_5 (\Delta \kappa_1+1)}{G_4 p^2} A_+(p)\\ \nn
&=\frac{2 z_0^2 k_+^4 \Delta k (\Delta \kappa_1+1)}{p^2 k_- } A_+(p)\\
&=\frac{2 k_+² \Delta k (\Delta \kappa_1+1)}{q^2 k_-} A_+(q)
\coloneqq \frac{η}{q²}A_+(q)\,,
\end{align}
where $Δk \coloneqq k_- - k_+$, and $η \coloneqq 2 (k_+²/k_-) \Delta k (\Delta \kappa_1+1)= 4 k_+^3(G_5/G_4)(\Delta \kappa_1+1)$. We will assume $k_- \sim k_+ \equiv k$ from now on.
Note that $z_c$ cancels out and there is only a dependence on the physical momentum $q$ on the dark bubble. This is, then, the missing condition. One can explicitly verify that this also solves (\ref{eq:secondjunct}) if we replace  $z_0$ with $z_c$ and take $z_c$ large, as it should by construction.

Using \cref{eq:firstjunction}, (\ref{eq:secondjunct}) and \cref{eq:Bcond} we can now fully determine all three functions up to the constant $\Delta \kappa_1$. In particular, we find 
\begin{multline} \label{eq: A_-}
    A_-(q)=\frac{16 \pi G_5 M(q)}{q} \times \\
    \times \frac{\eta I_2+q^2K_2^+}{\eta (I_2 K_1^-+I_1 K_2^-)+q^2(K_1^-K_2^+-K_1^+K_2^-)}\,,
\end{multline}
where $η$ is the constant of proportionality defined above. This expression is directly related to the gravitational potential given by
\begin{equation}\label{eq:potential}
V(\rho)=\sqrt{\frac{2}{\pi}}\int_0^\infty dq \frac{A_-(q) K_2^-\sin q \rho}{\rho}.
\end{equation}
In the limit of $q \rightarrow 0$ we find 
\begin{multline} \label{eq: A_-K_2}
  A_-(q) K_2^- \rightarrow -\frac{64 \pi G_5k^2 M(q)}{2 k² Δk -k \eta}\frac{k^2 }{q^2}=\frac{32 \pi G_5M(q)}{q^2 Δκ_1}\frac{k²}{Δk}\\
  =\frac{16 \pi G_4 M(q)}{q^2}.
\end{multline}
As we have seen, we need to choose $\Delta \kappa_1 =1$ in order for the hologram to match the physics of the dark bubble. We will come back to a comparison with the results for homogeneous radiation a bit later.

The potential given above is valid also in the hologram using the first junction condition, with an important difference. We need to replace $z_0$ with $z_c$ everywhere {\it except} in the expression for $\eta$ where $z_0$ should remain. As a consequence, the explicit $G_4$ in $\eta$ is effectively replaced by $(z_c^2/z_0^2) G_4$. This is in perfect agreement with our reasoning above, with the hologram reproducing the physics on the dark bubble with a re-scaled 4D Planck constant. What happens in practice is that the second junction condition on the dark bubble, as given by (\ref{eq:secondjunct}), is replaced by the subtracted extrinsic curvature at the cutoff. The term $A_-(p)$ is replaced by the subtraction. 

\subsection{Massive sources}

 For a point mass, we want the Fourier transform to be a $\delta$-function at the origin. This implies 
 \begin{equation}\label{eq:M4}
     M(p) =\frac{p}{4π\sqrt{2π}}M_4\,,
 \end{equation}
where $M_4$ is the mass on the brane defined such that
\begin{multline}
    \int_{\Omega_2}\sin θ\, dθ\, dϕ\int_0^∞ dr\, r²  \left(\frac{1}{r} \sqrt{\frac{2}{\pi}}\int_0^\infty dp \sin(p r) \frac{p}{4π\sqrt{2π}}\right) \\ = M_4\,.
\end{multline}
where we note that, using partial integration in the second step, we have
\begin{multline}
    \int_0^\infty dr r^2  \left(\frac{1}{r} \sqrt{\frac{2}{\pi}}\int_0^\infty dp \frac{p}{\sqrt{2π}} \sin (p r)\right) \\
    = \int_0^\infty dr r^2 \left( -\frac{1}{r
    }\frac{d}{dr}\delta (r) \right)
    =\int_0^\infty dr \delta (r) = 1.
\end{multline}
The energy momentum tensor becomes
\begin{equation}\label{eq:tmunu-mass}
T^a_b = 
\begin{pmatrix}
-M(p) & 0 & 0 & 0\\
0 & 0 & 0 & 0 \\
0 & 0 & 0 & 0 \\
0 & 0 & 0 & 0
\end{pmatrix}
=-\frac{M(p)}{3}\left(G_1+2G_2\right),
\end{equation}
where $G_i$ are the matrices defined in \cref{eq:G-matrices}.
Comparing this to the trace of \cref{eq:Sab-diff} we find
\begin{equation}
    F(p)=-\frac{k_- k_+ M(p)}{3 p^2}.
\end{equation}
As already observed in \cite{Padilla:2004mc}, it is difficult to solve the junction conditions when the brane is bent. The trick is to make a coordinate transformation in the bulk that makes the brane straight, and removes the non-zero $F(p)$. The calculation requires care, but the result is simple. In \cite{Banerjee:2020wix} the result in the $p \rightarrow 0$ limit was shown to be~\footnote{From here on, we absorb $z_0$ and use the physical 4D momentum $q$. In fact, this is a highly non-trivial check that the dark bubble reproduces relativistic gravity in line with Einstein gravity.}
\begin{equation} \label{eq: h small q}
    h_{ab}=-\frac{16 π G_4}{q^2} \left( T_{ab}-\frac{1}{2} T \eta_{ab} \right)\,.
\end{equation}
This has the right tensor structure for 4D gravity, i.e. $(1/2)Tη_{ab}$ instead of $(1/3)T η_{ab}$, but the wrong sign for a point particle of mass $M_4$. However, this is not the whole story. As shown in \cref{eq:ABsign}, gravitational backreaction (parametrized by the normalizable term $B_+(p)$) causes the non-normalizable term $A_+(p)$ (which is responsible for 4D gravity) to flip its sign. This crucial sign flip is what led to the value of $η$ in \cref{eq:Bcond}, and the solution for $A_-(p)$ in \cref{eq: A_-}. As explained above, this $η$ is different from the one in \cite{Banerjee:2020wix}.
This difference affects the coefficient of the  leading order in $q$, and qualitatively changes the  behavior at higher orders in an important way. Thus, we conclude that the full result to all orders $q$, with non-vanishing trace, must be given by
\begin{equation} \label{eq:darkmetric}
     h_{ab}=\frac{16 π G_4}{q^2}{\cal G}(q)\left( T_{ab}-\frac{1}{2} T \eta_{ab} \right),
\end{equation}
where
\begin{align} \label{eq:calg}
    {\cal G}(q)&=\frac{kG_5}{G_4} \times \\ \nn
    &\hspace{-2em}\times \dfrac{q^2(\frac{8k^3G_5}{G_4} I_2+q^2K_2^+)K_2^-}{k q\left[\frac{8k^3G_5}{G_4} (I_2 K_1^-+I_1K_2^+)+q^2(K_1^-K_2^+-K_1^+K_2^-)\right]}\nonumber  \\
    &\hspace{-2em}= \frac{ q^4K_2^2}{ 8k^4 +2q^3[q (K_2^2 -K_1^2)- 3 k K_1 K_2]} + {\cal O}(\Delta k).
\end{align}
To get the first line, we note that (\ref{eq: h small q}) should be compared with the limit taken in (\ref{eq: A_-K_2}), while the full expression needs (\ref{eq: A_-}).
In the second line, we have taken the limit $k_--k_+=\Delta k \rightarrow 0$, with $G_5/G_4 = Δk/(2k²)$, and kept the leading term. We have also used $I_2K_1^-+I_1K_2^+=I_2(q/k_+)K_1(q/k_-)+I_1(q/k_+)K_2^+(q/k_+)\rightarrow I_2(q/k)K_1(q/k)+I_1(q/k)K_2^+(q/k)=k/q$. 
to leading order in $\Delta k$. Note that ${\cal G}(q) \rightarrow 1$ when $q\rightarrow 0$ so that 4D general relativity is recovered. At large $q$, $I_i$ dominates,\footnote{$\lim_{q→∞}I_i(q/k_+)K_j(q/k_-) = \exp\left(\frac{Δk}{k_- k_+}\right) \frac{\sqrt{k_+ k+}}{2q}$, and $ \lim_{q→∞}I_i(q/k_+)K_j(q/k_-) = \frac{k_+}{2q}$. So only the $I_1 K_2^-$ and $I_2 K_1^-$ terms dominate in the limit.} $\eta$ (i.e. $\Delta \kappa_1$) cancels, and we find the following.
\begin{equation}
    {\cal G}(q) \rightarrow \frac{qG_5}{G_4}.
\end{equation}
This implies that 4D gravity is replaced by the much weaker 5D gravity such that the potential becomes (in terms of the mass on the brane, $M_4$, defined in \cref{eq:M4}):
\begin{equation}
  \frac{G_5 M_4}{\rho^2} \,\,\,{\rm for} \,\,\, \rho \rightarrow 0.
\end{equation}
At large distances, we have the expected expression for 4D Newtonian gravity, while at small $\rho$ we find that we are left with the much weaker 5D gravity. At intermediate distances, we have a non-trivial behavior that we will examine in the next section.

Note that if the expected Schwarzschild radius in usual 4D gravity is below the AdS$_5$ length $L$, then the theory will (as promised) be weakly coupled all the way down to $\rho\sim l_4$. This is because
\begin{equation}
    \frac{G_5 M_4}{l_4^2} \sim \frac{G_4 M_4 \Delta k}{k^2 l_4^2} <\frac{L \Delta k}{k^2 l_4^2} \sim \frac{L^2}{N l_4^2} \sim 1.
\end{equation}
Interestingly, this means that we have an example of a 5D C-metric in AdS. Instead of a black hole accelerated by a pulling string, our point mass is held at a constant radius in the AdS throat by the brane.\footnote{As far as we know, exact 5D C-metric solutions do not exist in the literature. Our solution is valid only for very tiny masses, where the 4D Schwarzschild radius would have been much smaller than the AdS radius, i.e, a 4D mass much smaller than roughly the mass of the moon.}

\subsection{Examining the gravitational potential}\label{sec:potential}

\begin{figure*}
	\centering
	\begin{subfigure}[t]{0.45\linewidth}
    \includegraphics[width=\linewidth]{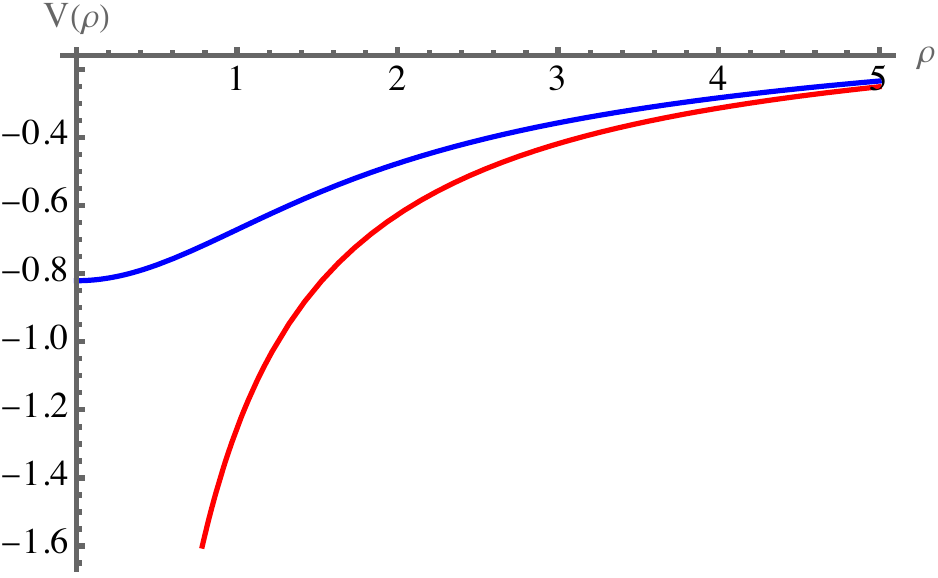}
    \caption{Gravitational potiential\label{fig:pot}}
    \end{subfigure}
    ~
    \begin{subfigure}[t]{0.45\linewidth}
    \includegraphics[width=\linewidth]{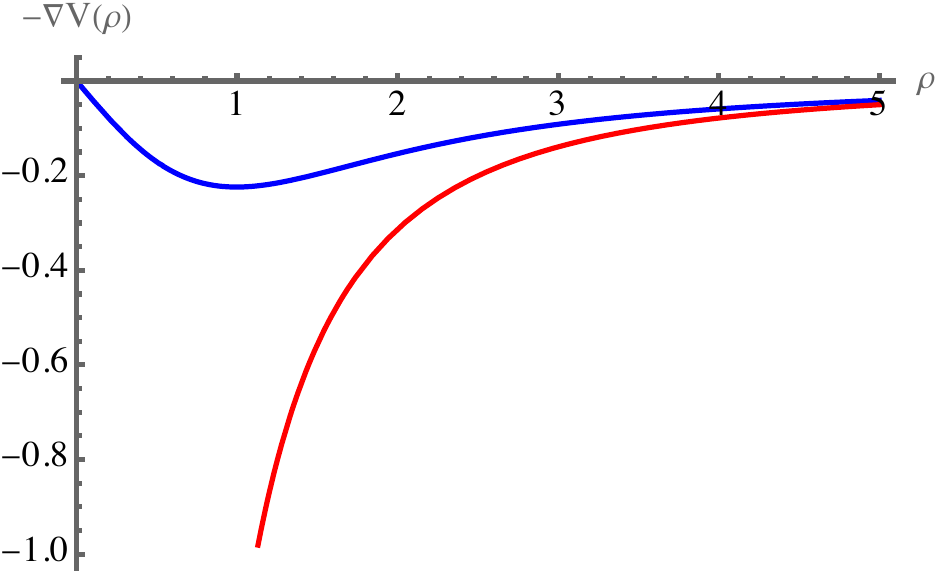}
    \caption{Gravitational force\label{fig:force}}
    \end{subfigure}
    \caption{The gravitational potential and the gravitational force (both in units of $G_4 M_4$) due to a particle of mass $M_4$, as a function of distance from the particle (in units of the AdS$_5$ length $L$). Lower curves correspond to ordinary 4D gravity, while the upper curves correspond to the the 4D universe on a dark bubble. Gravity on the dark bubble gets weaker at small distances.}
\end{figure*}

Let us now study the gravitational potential around a point mass in more detail.  It is enough to use the leading $\Delta k$ expression to cover the behavior of the potential from infinity all the way to the 4D Planck scale. Similar to \cref{eq:potential}, Fourier transforming the time-time component of the metric in \cref{eq:darkmetric} sourced by the point mass in \cref{eq:tmunu-mass}, we get the gravitational potential
\begin{multline}
   V(\rho)=\frac{8πG_4}{\rho}\sqrt{\frac{2}{\pi}}\int_0^\infty \frac{dq}{q^2}{\cal G}(q) M(q)\sin q\rho \\
   =\frac{8π G_4 M_4}{\rho}\frac{1}{4π\sqrt{2π}}\sqrt{\frac{2}{\pi}}\int_0^\infty \frac{dq}{q}{\cal G}(q) \sin q\rho ,  
\end{multline}
with ${\cal G}(q)$ given by \cref{eq:calg}. Note that we only need the leading term in $\Delta k$, i.e. the 4D gravity part. The first few terms at large $r$, are given by

\begin{equation}
    V(\rho )= G_4 M_4 \Bigg[ \frac{1}{\rho}-\frac{3L²}{2\rho^3}+\frac{3L^4\left(31-18\log (ρ/L)\right)}{\rho^5}+...\Bigg].
\end{equation}
The potential is depicted in \cref{fig:pot}, and the gravitational force is depicted in \cref{fig:force}, where we have used (\ref{eq:calg}), dropping the terms linear in $\Delta k$. We see how the potential flattens when the radius becomes as small as $L$, and the gravitational force is reduced.

We will soon move on to investigate the cosmological consequences of this unusual behavior of the gravitational force. But first, let us take a step back and interpret the results.

\section{Why the dark bubble might not have an EFT}\label{sec:why-no-eft}

Gravity in the dark bubble model is different from the one of Einstein gravity and its Newtonian limit. To find a language to describe the differences, we make contact with some general ways to modify gravity in the literature. In Newtonian gravity we can distinguish three kinds of masses. To see this, we relate the acceleration of particle number 1, to the gravitational field due to particle number 2 as
\begin{equation} 
m_{1i} \ddot r = -\frac{G_4 m_{1p} m_{2a}}{r^2}  .
\end{equation}
Here, $m_{1i}$ is the {\it inertial mass} of particle 1, $m_{1p}$ is the {\it passive gravitational mass} of particle 1, and $m_{2a}$ is the {\it active gravitational mass} of particle 2. The equivalence principle states that $m_i=m_p$, that is, the inertial and passive mass coincide. This is the essence of a theory in which gravity is captured by geometry. Einstein postulated this early on, but it was much more difficult to find the equations that determined how a mass sources the gravitational field. In the Newtonian limit this is the active gravitational mass. 

It is almost always assumed that $m_a=m_p$, and this is for good reasons. If $m_a/m_p$ is a universal constant different from one, we can always map back to $m_a=m_p$ by trivially changing $G_4$. But if they vary between different kinds of matter, the situation is different. In general, Newton's third law is broken and a system can self-accelerate even though no external forces are acting on the center of mass. This means that energy is not conserved. If no other forces act, other than the gravitational ones, one can restore energy conservation by re-defining energy in terms of the active gravitational mass rather than the passive one. If non-gravitational forces act, there is a true breakdown of Newton's third law. Such theories cannot be obtained from an action principle, and one can debate their consistency. See \cite{Fragkos:2025ieh} for some discussions and insights. 

In principle, the story can be generalized to the complete energy momentum tensor. We let $T_{ab}^{(p)}$ be the passive energy momentum tensor, while $T_{ab}^{(a)}$ be the active one. Let us now see how this can be a useful language to describe our results.

The passive energy momentum tensor $T_{ab}^{(p)}$ is the relevant one when gravity is ignored. The masses of elementary particles in particle physics are the inertial ones, which also coincide with the passive ones in order for the principle of equivalence to hold. This we identify with the energy momentum tensor representing the mass on top of the brane.

\begin{figure}
    \centering
    \includegraphics[width=\linewidth]{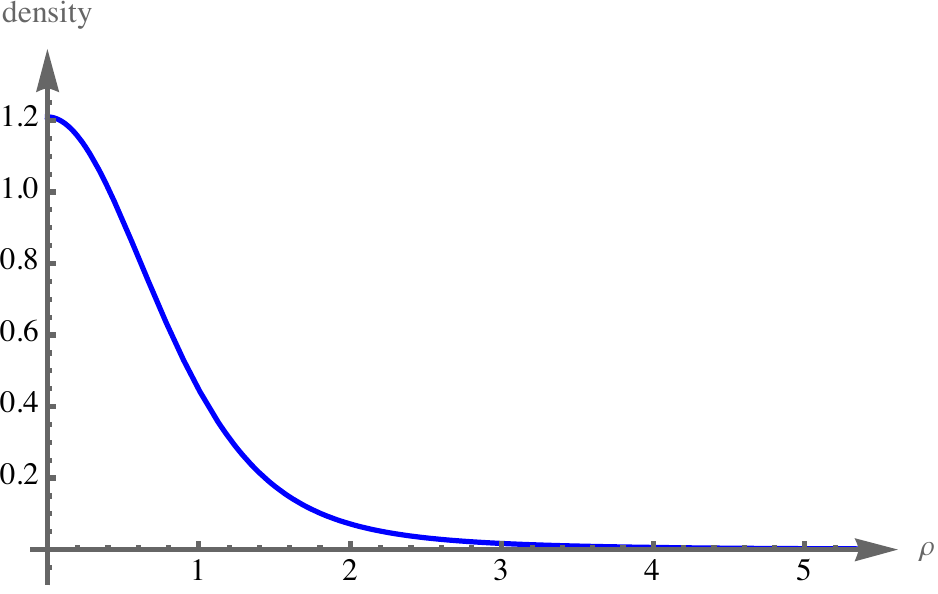}
    \caption{The active mass density for a point mass  as a function of radius in units of the AdS$_5$ radius $L$.}
    \label{fig:blob}
\end{figure}

The actual metric around a massive particle is not given by the passive energy momentum tensor but the active one given by $T_{ab}^{(a)}$. In Fourier space we see from  (\ref{eq:darkmetric}) that we have
\begin{equation} 
    T_{ab}^{(a)}={\cal G}(q)T_{ab}^{(p)}.
\end{equation}
From the gravitational point of view we find that any point mass with a mass less than $G_4 L/c^2$ will be smoothened out into a blob of radius $L$. In \cref{fig:blob} we see the density (given by the Fourier transform of $\mathcal{G}(q)M(q)$) as a function of the radius. Locally, this implies a difference between the passive and active gravitational masses, with equality restored at scales larger than $L$. Since the modification of the Newtonian force is universal, there is no breakdown of Newton's third law at any scale.

Using the distinction between passive and active energy momentum tensors to capture departures from general relativity is quite general, but not often used. The reason is simple. Assuming the existence of an effective action, deviations are captured using additional fields such as the dilaton, or extra terms involving the Riemann tensor and its derivatives. The situation with the dark bubble is less clear. The way gravity is induced is inherently 5 dimensional and whether a 4D action principle that can capture the modified gravity actually exists is far from clear. Basically, it would amount to introducing additional physical fields carrying energy momentum on which we can blame all deviations. If such an EFT does not exist, this could be the way the dark bubble evades the de Sitter swampland conjectures.

\section{Cosmological consequences}\label{sec:cosmology}

Let us now come to cosmology and a universe filled with homogeneous radiation. In what way does the change in the gravitational force at small scales come into play? 

We will start with the classical evolution. As we will see, the weakening of the gravitational force at small distances that we have observed for localized matter, translates into gravity becoming weaker at the high energy densities in the early universe. Effectively, this amounts to a reduction in the {\it active} energy density. This has dramatic consequences for the cosmology of the early universe.

We then address the cosmological origin of the dark bubble. As argued in \cite{Danielsson:2021tyb}, the nucleation of a dark bubble in AdS$_5$ is mathematically equivalent to Vilenkin's quantum cosmology, \cite{Vilenkin:1982de,Vilenkin:1984wp}. Contrary to the no-boundary proposal of Hartle and Hawking, where the universe appears from nothing, our universe appears through tunneling from something. While the purely 4D description is somewhat abstract, the dark bubble interpretation is intuitive. As we will see, weakening of the gravitational force at small distances will play a crucial role in the quantum cosmological origin of the universe.

Finally, we speculate on the observational consequences for the late time universe.

\subsection{Gravity gets weaker at high densities}

All we need to know about classical cosmology can be inferred from the junction conditions 
\begin{multline}
    \sigma + \rho_r=\frac{3}{8\pi G_5a} \Bigg( \sqrt{k_-^2 a^2+1+ \dot{a}^2}\\
    -\sqrt{k_+^2 a^2+1+ \dot{a}^2-\frac{2G_5 M_+}{a^2}}\Bigg) .
\end{multline}
We have put $a(\tau)=z_0(\tau)$ to make the notation more familiar. Note, though, that $a$ is not a dimensionless scale factor but has dimensions of length and is identical to the actual radius of the universe. We have assumed that the metric inside of the dark bubble is pure AdS, while it is AdS-Schwarzschild on the outside. In addition, we have allowed for radiation on top of the dark bubble in the spirit of \cref{sec:matter-on-the-brane}. It is convenient to parameterize the energy density as $\rho_r=3M_-/(8\pi k_-a^4)$. Furthermore, we have $\sigma = 3(k_--k_+)/(8\pi G_5)-\rho_{\Lambda_4}$, where $\Lambda_4$ is the 4D energy density from the cosmological constant. We now rearrange the junction condition to put it in the form of Friedmann's equation (for simplicity we ignore the late time cosmological constant),
\begin{multline}
    \dot{a}^2=-\Bigg[1+k_-^2 a^2-\frac{a^2}{4} \Bigg( k_--k_+ +\frac{G_5M_-}{k_- a^4}\\
    +(k_-+k_+)\frac{1+\frac{2G_5M_+}{(k_-+k_+)\Delta k a^4}}{1 +\frac{G_5M_-}{ k_-\Delta k a^4}} \Bigg)^2\Bigg]=-2V(a),
\end{multline}
which to lowest orders in $\Delta k$ can be written as 
\begin{equation} \label{eq:corrFriedman}
    \dot{a}^2=-\left[1+k^2 a^2-k^2 a^2\left(\frac{G_5M_-}{2k^2 a^4} +\frac{1+\frac{G_4M_+}{2k^3 a^4}}{1 +\frac{G_4M_-}{ 2k^3 a^4}}\right)^2\right].
\end{equation}
In the limit of small energy densities, $ρ_r \ll σ$, we find (up to sub-leading order in $1/a$),
\begin{equation}\label{eq:Friedmann-eq0}
\frac{\dot{a}^2}{a^2}=-\frac{1}{a^2}+\frac{G_4(M_+-M_-)}{ka^4}=-\frac{1}{a^2}+\frac{8\pi G_4}{3}\rho_r .
\end{equation}
provided that we make the choice $M_+=2M_-$, which is equivalent to the choices made for localized matter. This is just the Friedmann equation for a universe with radiation. As we will see, this choice is the unique one that allows the function ${\cal G}(q)$ to interpolate between $1$ and $qG_5/G_4$ without additional numerical coefficients, not only in the localized case, but also in cosmology.
\begin{figure}
    \centering
    \includegraphics[width=\linewidth]{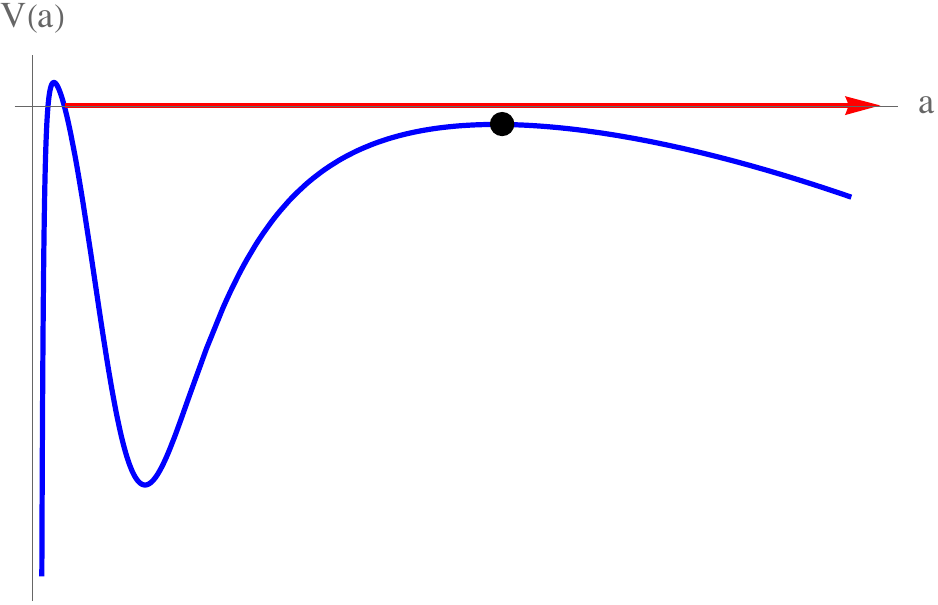}
    \caption{A schematic plot of effective potential for the scale factor, if the kinetic term as assumed to be $\dot{a}^2/2$. In order for all phases to be clearly visible, the figure is not to scale. The universe nucleates in front of a barrier close to the horizon of a 5D black hole in AdS. There is an early phase of accelerated expansion before entering into a decelerated, radiation dominated phase. The black dot marks the static Einstein universe. After passing over it, the universe enters into the late time accelerated phase.  }
    \label{fig:cosmicpot}
\end{figure}

Now let us see what happens if $a$ gets small and the energy densities large. At large $a$, the terms with $G_4$ completely dominate since $G_4 \gg kG_5$. However, we see from \cref{eq:corrFriedman} that the usual divergence of the energy density as $a\rightarrow 0$ is removed, and that in this limit the $G_4$ contribution is actually finite (although the divergence from the $G_5$ contribution remains).
This is nothing else than the cut off of 4D gravity at small scales that we have seen before. This happens when $G_4 M/(k^3 z^4) \sim 1$, or $\rho_r \sim 1/(G_4 L^2)$—that is, when the Hubble radius is of order $L$. This is in perfect agreement with what we have found in previous sections. Gravity gets weaker when the characteristic scale of physics approaches the 5D AdS-scale. Note that $1/(l_4^2 L^2)\sim 1/(Nl_4^4)\sim 1/(l_s^4)$, which corresponds to energy densities of order string scale.\footnote{It was observed already in \cite{Basile:2023tvh} that the Friedmann equations received corrections for such energy densities, but their ultimate cause was not identified.} Furthermore, we see that the Friedmann equation at even higher densities, will be dominated by the $G_5$ term that previously was subleading. Again, this is in line with the properties of the gravitational force outlined earlier in the paper.

\subsection{An inflationary phase}

Let us now try to understand the shape of the effective potential. At large $a$, if we take the cosmological constant into account,\footnote{Corresponding to the additional term $-ρ_{Λ_4}$ in $σ$ on the right hand side of \cref{eq:Friedmann-eq0}, that we otherwise ignore.} there is the familiar hill, with the top representing Einstein's static universe. At larger $a$ we find the present accelerated expansion. As we move to smaller $a$, the potential dives downwards as the energy density of the radiation increases. But as we have seen, this behavior is tamed, and the potential turns up again to reach another peak. See \cref{fig:cosmicpot}. It is this peak that governs the creation event. The 5D term will only kick in when $G_5M_-/(2k^2a^4)\sim 1$, or $\rho_r \sim 1/(G_5 L)\sim 1/l_4^4$, i.e., 4D Planckian densities.  This is much higher than the string densities where 4D gravity shuts itself off. Hence we can approximate the Friedmann equation at small $a$ as   \begin{equation}\label{eq: potpeak}
    H^2=-\left(\frac{1}{a^2}+k^2-k^2 \left(\frac{G_5M_-}{2k^2 a^4} +2\right)^2\right) ,
\end{equation}
where we have put $M_+=2M_-$. Actually, the term $G_5M_-/(2k^2a^4)$ can be ignored all the way back to 4D Planckian densities, where the scale factor is reduced by a further factor $N^{1/4}\sim10^{15}$ compared to the stringy densities. Over all these orders of magnitude the effective Friedmann equation simply becomes
\begin{equation}
    H^2=3k^2\,.
\end{equation}
In other words and quite surprisingly, we find an inflationary phase, where the Hubble constant remains at order $H\sim 1/L$ over more than 30 e-foldings. Note that the inflationary period is not due to the presence of a constant vacuum energy. The universe is still dominated by radiation, but the gravitational response is altered due to a reduction in the active energy density.

Actually, the horizon problem is already solved in dark bubble cosmology, even without the inflationary phase, through the initial nucleated bubble, which starts out at rest and completely within its causal horizon.\footnote{This is similar to string gas cosmology~\cite{Brandenberger:1988aj,Brandenberger:2008nx}.} When the bubble first starts to move, the densities are Planckian and 5D gravity plays a role.\footnote{Note the parallel with the point mass where we saw that 5D gravity became important at the 4D Planck scale.}  As the densities reduce, 5D gravity becomes weak, and the inflationary phase takes over. It is not terminated until the energy densities are reduced to string scale and 4D gravity becomes relevant. The fact that the Hubble constant remains at the constant value $\sqrt{3}k$ for such a long period suggests that the dark bubble model can reproduce a scale invariant spectrum of cosmological perturbations. We will return to this possibility in later work.

\subsection{Nucleation}

Planckian densities also coincide with the approach towards the horizon of the 5D AdS black hole, where 5D gravity takes over. We now get an intricate interplay between the various scales. First, we note from (\ref{eq:corrFriedman}) that $M_+>M_-$ is a necessary condition for a peak with a potential that is decreasing as $a$ is increased and we move away from the peak. Second, as a consequence of this, the peak must sit at some $a_p=a<1/k=L$, if we want $V(z_p)>0$. This implies that the horizon scale must be smaller than $L$ as well, and that the black hole needs to have a mass $M<L^2/G_5$, and cannot be a fully developed AdS-black hole. This is a bit surprising and has important observational consequences that we will come back to in the next section.

In \cite{Danielsson:2021tyb} the creation of an empty universe with nothing other than a positive cosmological constant was studied. Here, we will consider the semi-realistic universe filled with radiation as described above. This means that the nucleation is catalyzed by an initial 5D black hole, similar to the 4D analysis of \cite{Gregory:2013hja}.
If we want to quantize the system, we need to implement a Hamiltonian formalism and find the conjugate momentum. The result found in \cite{Danielsson:2021tyb} is
\begin{align} \label{eq:H}
    H&=2\pi^2  a^3 \left(\sigma - \frac{3(\beta_--\beta_+)}{8\pi G_5a}\right) \nonumber \\ \nn
    &= 2\pi^2  a^3 \sigma-\frac{6\pi^2a^2}{8\pi G_5}\left(\alpha_-+\alpha_+-2\sqrt{\alpha_-\alpha_+}\cosh{\frac{4G_5 p}{3\pi a^2}}\right)^{1/2}\\
    &=0,
\end{align}
with respect to proper time $\tau$, with 
\begin{equation}\label{eq:cosheq}
\cosh{\frac{4G_5 p}{3\pi a^2}}=\frac{\beta_-\beta_+-\dot{a}^2}{ \sqrt{\alpha_-\alpha_+}}\,,
\end{equation}
where $\beta_\pm=\sqrt{\alpha_\pm+\dot{a}^2}$. In our case  $\alpha_+=k_+^2a^2+1-2G_5 M_+/a^2$, while $\alpha_-=k_+^2a^2+1$. We have also added radiation on the dark bubble so that $\sigma \rightarrow \sigma + \rho_r=\sigma+3M_-/(8\pi k_-a^4)$, with $M_+=2M_-$.
For simplicity we have put the lapse function, which does not affect any physical results, to one. Denoting $Hd\tau={\cal H} dt$, with\begin{equation}
    d\tau = \frac{1}{2}\left(\beta_-+\beta_+\right) dt ,
\end{equation}
we recover the bulk Hamiltonian ${\cal H}$ obtained in (\ref{eq: Hbulk}).

This has two consequences. First, the Hamiltonian is not quadratic in the momentum unless $p$ is small. Second, even in this limit, the effective potential with a $p^2$ kinetic term will have a non-trivial $a$-dependent coefficient, compared to the one of the previous section. To be precise, expanding \cref{eq:cosheq} to leading order in $\dot{a}$, we find, for small $p$, 
\begin{equation}
    p^2\sim\frac{9\pi^2}{16G_5^2} \frac{a^4\dot{a}^2\left(\sqrt{\alpha_-}-\sqrt{\alpha_+}\right)^2}{\alpha_-\alpha_+}.
\end{equation}
In the limit $k_\pm a \gg 1$ we find that $p^2\sim a^2 \dot{a}^2$.  This is the potential that was plotted and used in \cite{Danielsson:2021tyb}. The corrected version of the potential in \cref{fig:cosmicpot} looks qualitatively the same, and we do not make a separate plot. Whether or not the Hamiltonian remains in the quadratic regime, depends on how close the turning points are to each other and to the horizon.
 
\begin{figure}%
    \centering
    \includegraphics[width=\linewidth]{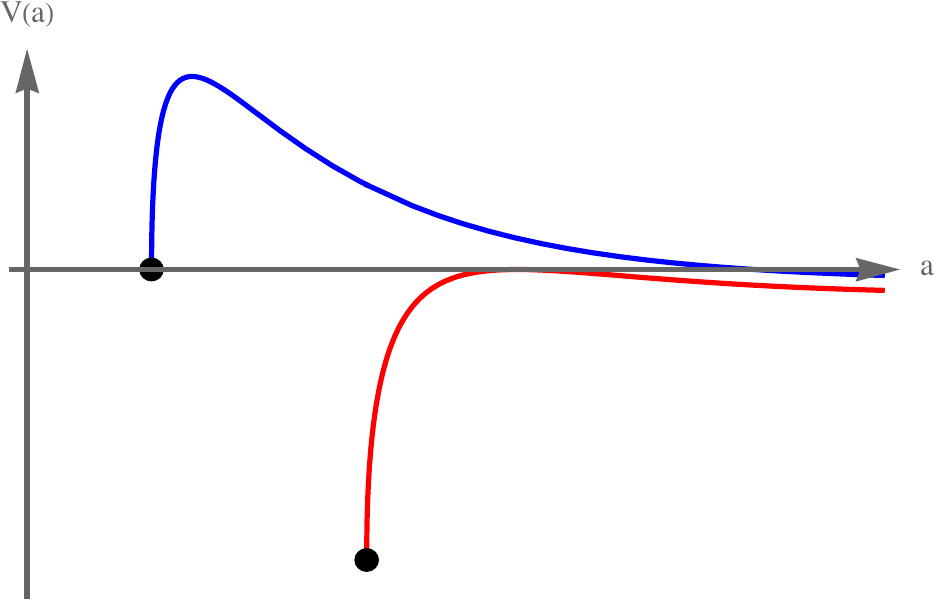}
    \caption{The effective potential near the top of the potential, where physics is dominated by 5D gravity, is plotted for the two extreme values for $M_-$. The upper curve corresponds to the lower limit of $M_-$, where the shell nucleates at the horizon and tunnels through a barrier and escapes. The lower curve corresponds to the higher limit of $M_-$, which allows for a balancing solution at the top of the potential. The black dots represents the position of the horizon in the two cases.}
    \label{fig:tunneling}
\end{figure}

The endpoints of the barrier are determined by putting $p=0$ in \cref{eq:H}, yielding 
\begin{equation}
    \frac{G_5M_-}{k_-a^3}-\sqrt{\alpha_-}+\sqrt{\alpha_+} = 0,
\end{equation} 
where we have used that $\rho_r \gg \sigma$ close to the peak. If choose to put the inner edge of the barrier at the Schwarzschild radius, such that $k^2 a_i^2+1-2G_5M_+/a_i^2=0$, we find, imposing $M_+=2M_-$, that $M_-=4/(225 k^2 G_5)\sim 0.0178/(k^2 G_5)$. Lower values will push the barrier inside the horizon and are therefore excluded. Higher values  will create a well surrounding the potential similar to the cosmic egg of \cite{Hertog:2021jyd}. If $M_->(13\sqrt{13}-46)/(24 k^2 G_5)\sim 0.0363/(k^2 G_5)$, the barrier will be completely removed. In \cref{fig:tunneling} we have made a plot of the effective potential (defined as the Hamiltonian evaluated at $\dot{a}=0$) for the critical values of $M_-$ that we have derived.\footnote{It should be possible to calculate the   actual tunneling amplitudes for various cases along the lines of \cite{Gregory:2013hja}. We will return to this important task in future work.}

\subsection{Late time observational consequences}

Let us now check if our estimated mass of the original 5D black hole is consistent with the present universe. In particular, we will derive a constraint on its radius. According to the dark bubble model, the density of radiation is given by $\rho_r = 3 \chi/(8 \pi k^3G_5a^4)$, where $a$ is the current radius of the universe, and the constant $\chi$ is such that $0.0178<\chi<0.0363$, if the mass of the 5D black hole is in the interval argued earlier. To translate this into a numerical value, we need several results from \cite{Danielsson:2023alz}. These are $1/(k^3G_5)=2N^2/\pi$, which follows from standard rules of compactification\footnote{As observed in \cite{Danielsson:2024frw} this is not limited to $S^5$. The same relation is obtained for any manifold where a D3-brane has critical tension.} in AdS$_5 \times S^5$, as well as $1/(k^2G_4)=2N/(3\pi)$, which makes use of the second junction condition of the dark bubble, assuming that the bubble consists of a single D-brane. We also need $\rho_\Lambda=4/(3\pi g_s L^4)$, which relates the 4D cosmological constant to $\alpha^\prime$-corrections with $g_s$ as the string coupling.

With these relations we derive $1/a_H^2=(8\pi G_4/3)\rho_\Lambda=8\pi^2/(g_s G_4 N^2)$ and $\rho_r=(16\pi\chi/g_s)(a_H^4/a^4)\rho_\Lambda$. Finally, we need the relation $g_s=(2/3)\alpha_{\textrm{\scshape{em}}}$, where $\alpha_{\textrm{\scshape{em}}} \approx 1/137$ is the fine structure constant. This was derived in  \cite{Danielsson:2023alz} from matching with an extremal Reissner-Nordström black hole realized as a stretched string. The result was corroborated in \cite{Danielsson:2024frw}, from studying charged Nariai black holes on the dark bubble. Thus we conclude that 
\begin{equation}
    \Omega_c=\frac{a_H^2}{a^2}=\sqrt{\frac{\alpha_{\textrm{\scshape{em}}}\Omega_r}{24\pi\chi\Omega_\Lambda}}\,,
\end{equation}
where $\Omega_i =\rho_i/\rho_{\rm crit}$.
Since we have not been able to measure any positive curvature of the universe, we expect $a \gg a_H$.  But what is the relevant value of $\rho_r$, i.e. $\Omega_r$? The fact that some small fraction of the energy in the early universe was converted into massive particles that now dominate the energy density of the universe (except for the cosmological constant), is irrelevant. The energy density we should focus on is still the one of radiation today, which is $\Omega_r \approx 5 \times 10^{-5}$, together with $\Omega_\Lambda \approx 0.7$. Using $\chi$ in the narrow interval given earlier, this suggests $\Omega_c\approx  5\times10^{-4}$,  which is below current limits. The conclusion is that the amount of radiation in our universe can be accounted for by a 5D black hole of the suggested size, even though it is microscopic. \emph{The prediction is that a positive value of the curvature is soon to be discovered.} In some sense, this result can be viewed as an explanation for the why-now-problem (or coincidence problem) in cosmology—why is $\Omega_{\textrm{matter}}$ and $\Omega_Λ$ of the same order of magnitude today, even though $\Omega_{\textrm{matter}} \sim a^{-3}$, while $Λ$ remains constant in time.

Another important quantity to check is the entropy. The entropy of the initial 5D black hole is of order $L^3/G_5 \sim N^2$, and can never decrease even after a bubble has nucleated. When this happens, the entropy needs to be carried by matter on the brane. This excludes the nucleation of an empty brane, and forces a non-zero value of $\rho_r$. Initially, the entropy will be carried by $N^2$ degrees of freedom, each at temperature $T\sim 1/L$, which is the temperature of the black hole. This accounts for the total mass through $N^2 T^4 \times L^3\sim N^{3/2} \sim L^2/G_5 \sim M$, and makes sure that the total entropy is $N^2 T^3 \times L^3 \sim S$. As the universe expands, we expect the number of degrees of freedom of radiation to be reduced all the way down to ${\cal O}(1)$ with only electromagnetic light remaining in the present universe. As this happens, the temperature will increase relative to the usual $1/a$ decay. This is similar to reheating in the inflationary paradigm.

For this change in the number of degrees of freedom to be possible, we need to keep track of where the entropy goes. If there are order one massless degrees of freedom, temperature can be estimated as $T^4 \sim N²/a^4$, and the total entropy carried by the gas is then $T^3 a^3 \sim N^{3/2}$. This is just a fraction of the total entropy $N^2$, and only captures entropy locally stored in the gas. The majority of the entropy is stored on scales larger than the Hubble scale, and is captured by the cosmological horizon with $a_H^2/l_4^2 \sim N^2$. The relation that makes it possible to match the entropy of the 3D horizon of the 5D black hole to that of the 2D de Sitter horizon in 4D is the following.
\begin{equation}
    L^3/G_5 \sim a_H^2/G_4 \sim N^2.
\end{equation}
It is only because of the to the dark bubble's unique property $G_5 \ll LG_4$ that this is possible.
It is also intriguing that the high entropy observed in the early universe in the form of radiation can be traced back to (a fraction of) the entropy of an initial 5D black hole. One also notes that the entropy during the inflationary phase is set by a horizon scale of order $L$, and a 4D Newton's constant given by $k G_5$. The entropy is again $L^2/(k G_5) \sim L^3/G_5 \sim N^2$. That is, most of the entropy is on scales larger than the horizon already at this stage.

It is natural to ask how this picture relates to swampland criteria, in particular the de Sitter and distance conjectures. In our case, accelerated expansion does not arise from a four-dimensional scalar potential, but from the higher-dimensional geometry and junction conditions. The effective four-dimensional description should therefore be treated with some care.

As discussed in Section III, the model may not admit a standard EFT valid over large field ranges. This is in line with the distance conjecture, where new light states are expected to appear and invalidate the low-energy description. The regime where gravity weakens can be viewed as approaching such a boundary.

From this perspective, the usual tension with the de Sitter conjecture may be avoided, since the construction is intrinsically higher-dimensional. A more precise connection would require an explicit string embedding, which we leave for future work.

\section{What about black holes in 4D?} \label{sec:blackholes}

The analysis of the paper shows that gravity is cutoff at distances smaller than $L$, so that no black holes with a Schwarzschild radius smaller than $L$ will form. Using the preferred value $L \sim 10^{-5}$m we note that this implies that there are no black holes with a mass smaller than the mass of the moon. It is tempting to speculate on the fate of black holes with a larger mass. Since gravity is likely to be unmodified at larger distances, we might naturally expect to find some kind of uplifted versions of black holes from the 5D point of view. However, there are reasons to expect something much more interesting, inspired by black shells acting as black hole mimickers, \cite{Danielsson:2017riq}.

According to the black shell proposal, black holes are replaced by bubbles of AdS-space contained by nucleated shells on top of which we have high entropy stringy matter. The shells nucleate through a phase transition where the cosmological constant drops down to a large and negative value. In empty space this process is extremely rare, that is why the universe still exists as we know it, but due to the enormous phase space of the gas on top of the shell it becomes inevitable at gravitational collapse. Thus, black holes are always replaced by black shells. The argument for why this could fit into the results of the present paper is as follows.

\begin{figure}%
    \centering
    \includegraphics[width=\linewidth]{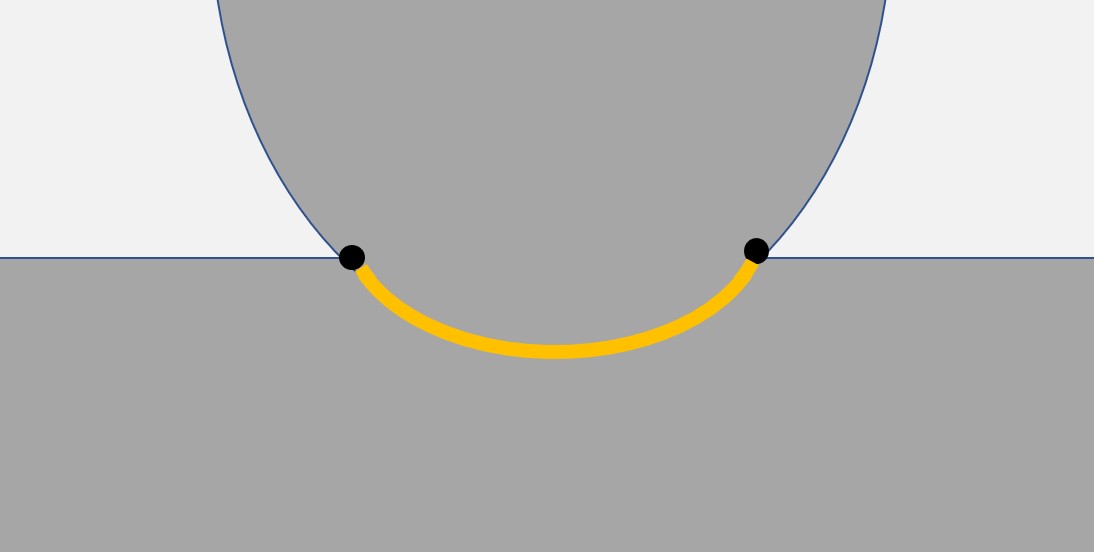}
    \caption{A higher dimensional picture of a black shell. The two black dots is a section through the shell. The yellow line is the interior AdS$_4$. }
    \label{fig:blackshell}
\end{figure}

What prevents the existence of the small black holes in the dark bubble model is that gravity is shut off at small distances. As we have seen, its strength reduces as $G_4=2k^2 G_5/Δk \rightarrow k G_5$. In our analysis this phenomenon only happens over a small region of size $L$, and we would expect the effect to be completely irrelevant for large black holes. But what if the nucleation of a black shell can be uplifted to 5D and incorporated in the dark bubble model? Imagine that a second dark bubble nucleates in the 5D bulk when a large black hole threatens to form in 4D. It then expands and hits the cosmological dark bubble as in \cref{fig:blackshell}. From the 4D point of view this is interpreted as the creation of a black shell. The region where the two branes join will form the interior of the black shell. Within the black shell, the 5D junction conditions will be those of two insides, just as in the case of the Randall-Sundrum model \cite{Randall:1999vf,Randall:1999ee}. Instead of the dark bubble relation $G_4=2k_-k_+ G_5/Δk$, the strength of the interior 4D force of gravity within the shell will be $k G_5$. That is, gravity is again shut off but in a region set by the size of the shell rather than just $L$. We find this to be a tantalizing clue to how black holes could be replaced by black shells in the dark bubble model. This effect also has immediate consequences for the black shell model and the need to protect against the collapse of matter penetrating the interior of the black shell. If gravity is weak inside of the black shell, this will not happen. See \cite{Giri:2024cks} for some related considerations. We will return to the black shells on dark bubbles in future work.

It is also interesting to consider the weak gravity conjecture. In our setup, gravity becomes weaker below the AdS scale, which may modify the balance between gravitational and gauge forces.

This suggests that the usual picture of extremal black hole decay could be altered, or realized differently, in the presence of the dark bubble. However, we do not have a concrete model of charged states in this background, and a quantitative analysis is left for future work.

A related issue is the stability of the dark bubble. While the solution follows from consistent junction conditions, this does not guarantee dynamical stability.

Possible concerns include a stabilization mechanism and the behavior of perturbations. A proper analysis would require studying fluctuations of the coupled bulk–brane system, which we leave for future work. We note, however, that the cosmological solutions considered here do not exhibit any immediate pathologies.

\section{Conclusion}\label{sec:conclusion}

In this paper, we have explored the gravitational field of localized 4D matter sources in the dark bubble model. We have considered matter fields confined to the dark bubble affecting the bulk geometry and the extrinsic curvature of the dark bubble brane. Crucial to our construction is the presence of non-normalizable modes associated with mixed boundary conditions at asymptotic infinity in the AdS-throat. Contrary to usual AdS/CFT holography, the hologram contains a theory {\it with} gravity. This theory is a conformally rescaled version of the theory on the dark bubble. 

In the effective 4D theory we find that gravity is essentially shut off at distances smaller than the AdS-scale $L$ of the higher dimension. As we have argued, this fundamental length scale predicted by the dark bubble model, is of order $10^{-5}$ m, given the measured value of the cosmological constant. This suggests experimental opportunities to test the dark bubble model in the laboratory.

We also re-examined cosmology in the presence of radiation and a positive cosmological constant. We discovered that the reduction in the force of gravity is present there as well. When densities are reached where the Hubble radius is as small as the AdS-radius, 4D gravity effectively shuts off. As we move further back in time, the Hubble constant remains at $H \sim 1/L$ over more than thirty e-folds, even though the energy density of radiation keeps growing. This effectively implements inflation and suggests that a scale invariant spectrum could be an automatic consequence of dark bubble cosmology. We will return to a detailed study of possible mechanisms in the future. 

We also saw how the theory predicts values of the amount of radiation in the present universe comparable with what is actually measured, and explains the large entropy in our universe as coming from the initial 5D black hole that catalyzed the nucleation of the accelerating dark bubble that we now live on.

\section*{Acknowledgements}

 We thank Souvik Banerjee, Vincent Van Hemelryck, and Daniel Panizo for useful discussions. Support from Kungliga Fysiografiska sällskapet i Lund is acknowledged. The work of S.G. was conducted with funding awarded by the Swedish Research Council grant VR 2022-06157.

\bibliography{references}

\end{document}